\documentclass{aa}
\usepackage{txfonts}
\usepackage{graphicx}
\usepackage{natbib}
\bibpunct{(}{)}{;}{a}{}{,}

\defcitealias{lam05}{LAM05}
\defcitealias{bc03}{BC03}

\begin{document}
\title{The mass function of young star clusters in spiral galaxies}

\author{S{\o}ren S. Larsen}
\institute{Astronomical Institute, University of Utrecht, Princetonplein 5,
  NL-3584 CC, Utrecht, The Netherlands}

\offprints{S.\ S.\ Larsen, \email{S.S.Larsen@uu.nl}}

\date{\today}

\abstract
{}
{
The initial cluster mass function (ICMF) in spiral discs is constrained and compared with data for old globular clusters and young clusters in starbursts.
}
{
For a given absolute magnitude, the cluster age distribution depends on the ICMF. Here, the behaviour of the median age-magnitude relation is analysed in detail for Schechter ICMFs with various cut-off masses, \mbox{$M_c$}. The calculated relations are compared with observations of the brightest clusters in spiral galaxies. Schechter functions are also fitted directly to observed mass functions (MFs).
} 
{
A single Schechter ICMF with an $\mbox{$M_c$}$ of a few times 10$^5\mbox{$M_{\odot}$}$ can reproduce the observed ages and luminosities of the brightest (and \mbox{$5^{\rm th}$} brightest) clusters in the spirals if disruption of optically visible clusters is dominated by relatively slow secular evolution. A Schechter function fit to the combined cluster MF for all spirals in the sample yields $\mbox{$M_c$} = (2.1\pm0.4)\times10^5\mbox{$M_{\odot}$}$. The MFs in cluster-poor and cluster-rich spirals are statistically indistinguishable. An $\mbox{$M_c$}=2.1\times10^5\mbox{$M_{\odot}$}$ Schechter function also fits the MF of young clusters in the Large Magellanic Cloud. If the same ICMF applies in the Milky Way, a bound cluster with $M>10^5\mbox{$M_{\odot}$}$ will form about once every $10^7$ years, while an $M>10^6\mbox{$M_{\odot}$}$ cluster will form only once every 50 Gyr. Luminosity functions (LFs) of model cluster populations drawn from an $\mbox{$M_c$}=2.1\times10^5\mbox{$M_{\odot}$}$ Schechter ICMF generally agree with LFs observed in spiral galaxies.
}
{
The ICMF in present-day spiral discs can be modelled as a Schechter function with $\mbox{$M_c$}\approx2\times10^5\mbox{$M_{\odot}$}$.  However, the presence of significant numbers of $M>10^6\mbox{$M_{\odot}$}$ (and even $M>10^7\mbox{$M_{\odot}$}$) clusters in some starburst galaxies makes it unlikely that the \mbox{$M_c$} value derived for spirals is universal. In high-pressure environments, such as those created by complex gas kinematics and feedback in mergers, \mbox{$M_c$} can shift to higher masses than in quiescent discs.
}


\titlerunning{The mass function of star clusters}
\maketitle

\noindent \emph{This is a revised version of Larsen (2009:\ A\&A 494, 539) that includes corrections in Larsen (2009:\ A\&A 503, 467).}

\section{Introduction}

Star clusters connect many branches of astrophysics. Classically, they have been important test benches for models of stellar evolution \citep{mm81}. A large fraction of all star formation occurs in clusters \citep{ll03}, and massive, long-lived globular clusters (GCs) are potentially useful tracers of stellar populations in their parent galaxies \citep{bs06}. It is therefore of considerable interest to understand how the properties of star clusters are related to those of their parent galaxy, and in particular whether the formation (and survival) of GC-like objects requires special conditions. Three important properties of cluster populations which bear on this issue can, to some extent, be studied separately: 1) the initial cluster mass function (ICMF), 2) the formation efficiency of (bound) star clusters \citep{lr00,bastian08,pf08}, and 3) dynamical evolution and cluster disruption, addressed both theoretically and observationally in many studies \citep{vz03,bm03,lam05,mf08,gb08,whit07}. All three may, in principle, depend on time and location within a galaxy. This paper is concerned with the mass function, primarily its high-mass end and the possible existence of an upper cut-off mass. 

Highly luminous, compact star clusters are observed to form in on-going starbursts such as those associated with major gas-rich mergers and, in smaller numbers, in many other external galaxies with elevated star formation rates \citep[e.g.,][]{lar04}. However, conversion of observed cluster luminosities to masses requires knowledge of the mass-to-light ratios which, in turn, are strongly age-dependent. Dynamical mass estimates via integrated-light measurements of velocity dispersions have been obtained for only a limited number of clusters, but have confirmed that young clusters with masses above $10^5 M_\odot$ (and, in some cases, $10^6 \mbox{$M_{\odot}$}$) are quite common in external star-forming galaxies \citep{hf96,lar01,lbh04,lr04,mg07,sg01,dp07}. On the other hand, most young star clusters known in the Milky Way are relatively sparse with masses of $10^2$ - $10^4 M_\odot$. This apparent contrast between the properties of Milky Way open clusters and the luminous star clusters encountered in some external galaxies raises important questions about the universality of the cluster formation process. Do the most massive clusters form preferentially in starburst environments for some physical reason, or do they just happen to be statistically more likely to form there because the total numbers of stars and clusters formed are very large? 

Most studies of the statistical properties of star cluster populations have concentrated on \emph{luminosity} functions (LFs), which are more readily derived for large samples of clusters than mass functions (MFs). There is a roughly linear relation between the luminosity of the brightest cluster in a galaxy and the total number of clusters, as  expected if luminosities are drawn at random from a power-law distribution with slope $\approx-2$ \citep{bhe02,lar02,weidner04,whit03,whit07}. This implies that if the LF has a physical upper limit, no cluster system studied to date is sufficiently rich to sample the LF up to that limit, and the (young) cluster populations in most galaxies might in fact be drawn from very similar underlying LFs. 

Clearly, the MF would be a physically more interesting property to study than the LF.  The LF generally 
depends on both the mass- and age distribution of the cluster sample in question \citep[e.g.][]{fall06}, and since the latter is often unknown, the observed LFs offer only limited insight into the underlying MFs.  If the goal is to obtain information about the \emph{initial} MF, disruption effects must also be taken into account.  In the small number of cases where MFs have been determined, they are generally found to be consistent with power-laws $dN/dM \propto M^\alpha$ with a slope $\alpha \approx -2$ 
\citep{bik03,deg03,mg07,zf99,dow08}. However, the mass ranges differ from one study to another (e.g.\ $2.5\times10^3 < M/\mbox{$M_{\odot}$} < 5\times10^4$ for the Bik et al.\ study of \object{NGC~5194}, and $10^4 < M/\mbox{$M_{\odot}$} < 10^6$ for \object{the Antennae} clusters studied by Zhang \& Fall), so it is not clear that the similarity of the slopes derived for different galaxies actually implies a universal ICMF. 

Even if the MF can typically be fit by a power-law, this might only apply over a limited mass range. One potential difficulty is the large number of very massive clusters predicted in the \object{Milky Way} for a uniform power-law MF. For a star formation rate in bound star clusters of $5.2\times10^{-10} \mbox{$M_{\odot}$}$ yr$^{-1}$ pc$^{-2}$ in the Solar neighbourhood \citep[][hereafter LAM05]{lam05}, the Milky Way disc would have formed about 100 clusters with $M > 10^6 \mbox{$M_{\odot}$}$ over a 10 Gyr lifetime, assuming that the cluster formation rate was constant in the past (probably a conservative assumption; e.g.\ \citealt{fuchs08}). These are not observed. Since the completeness of existing surveys of young Milky Way clusters remains extremely poorly quantified it cannot be excluded that some remain hidden behind massive extinction, and others may have dissolved (Sect.~\ref{sec:disrupt}). Another possibility is that they simply never formed, or at least in smaller numbers than predicted by extrapolation of a pure power-law MF.

An upper limit to the ICMF need not be a strict cut-off, but could also take the form of an exponential decline above some characteristic mass \mbox{$M_c$}, i.e.\ a \citet{sch76} function:
\begin{equation}
  \frac{dN}{dM} \propto (M/\mbox{$M_c$})^\alpha \exp \left(-M/\mbox{$M_c$}\right),
  \label{eq:schechter}
\end{equation}
According to \citet{bastian08}, the relation between the luminosity of the brightest cluster in a galaxy and the star formation rate (SFR) can be reproduced if clusters in all galaxies are drawn from a Schechter function with $\mbox{$M_c$} = (1-5)\times10^6\mbox{$M_{\odot}$}$. More indirect evidence of a truncation or steepening of the MF in young cluster systems comes from the detection of a bend in the LF of clusters in \object{NGC~5194} (M51), \object{NGC~6946} and the Antennae galaxies (NGC 4038/39), which is consistent with a power-law MF truncated at a mass of $5\times10^5 - 10^6\mbox{$M_{\odot}$}$ \citep{zf99,gieles06}.

The aim of this paper is to examine, for a larger sample of ``normal'' spiral galaxies, the behaviour of the cluster MF at its high-mass end and constrain the upper mass limit, $\mbox{$M_c$}$. The most direct way to do this would perhaps be to simply obtain luminosities and ages for a large sample of clusters in each galaxy and converting these to masses, ideally using \emph{Hubble Space Telescope} (HST) imaging to obtain clean cluster samples. Although a large amount of data for nearby galaxies have by now accumulated in the HST archive, many datasets suffer from two basic shortcomings: 1) they lack (deep) imaging in a (roughly) $U$-band equivalent filter, which is essential for age-dating of the clusters, and 2) they cover only a small fraction of the galaxy. The latter limitation is particularly problematic when studying the most massive clusters, which tend to be rare. The only spiral which is sufficiently nearby to take advantage of HST's imaging resolution for cluster identification and is fully covered by multi-passband HST imaging including the ultraviolet is NGC~5194 \citep{scheep07,hl08}, which is arguably not a ``normal'' spiral due to its on-going interaction with \object{NGC~5195}. 

Instead of studying the complete MF in detail for each individual galaxy, we begin by taking an approach similar to that adopted in several previous studies of the LF: from the correlation between the magnitude of the \emph{brightest} cluster and some estimate of the \emph{total} cluster population, one can test whether the data are consistent with random sampling from a single, universal LF. Here, we go one step further and also attempt to draw conclusions about the MF from observations of the brightest (and \mbox{$5^{\rm th}$} brightest) clusters.  As will become clear below, the crucial extra information that makes this possible are the cluster \emph{ages}, but because these are needed only for the brightest  clusters, ground-based data are generally adequate. 

The general outline of the paper is as follows: In Sect.\ \ref{sec:montecarlo} we determine the relation between the median age and luminosity of the brightest and \mbox{$5^{\rm th}$} brightest clusters (or, in fact, any sample with magnitudes in a fixed, narrow range) for a given sample size, cut-off mass $\mbox{$M_c$}$, and age distribution.  This statistical age-luminosity relation should not be confused with the general fading  of an individual star cluster that is the result of stellar evolution and cluster dissolution. The (log(age), $\mathcal{M}_V$) predictions are compared (in Sect.\ \ref{subsec:data}) with previously published ground-based data for clusters in a sample of nearby spiral galaxies in order to put constraints on \mbox{$M_c$}.  In Sect.\ \ref{sec:mfs}, a direct Schechter function fit to observed MFs is performed and yields $\mbox{$M_c$}=(2.1\pm0.4)\times10^5\mbox{$M_{\odot}$}$. In Sect.~\ref{sec:lfs} the properties of the LF for a cluster sample drawn from an $\mbox{$M_c$}=2.1\times10^5\mbox{$M_{\odot}$}$ Schechter MF are compared with observed LFs. In Sect. \ref{sec:elsewhere} the implications of an $\mbox{$M_c$}=2.1\times10^5\mbox{$M_{\odot}$}$ Schechter MF are discussed in the context of other cluster systems, and it is concluded that while such a MF is consistent with existing data for Milky Way and \object{Large Magellanic Cloud} clusters, it is unlikely to be universal. Finally, in Sect.\ \ref{sec:discussion} the results are discussed in the context of host galaxy properties and current views on star/cluster formation.

\section{Constraining the upper limit of the MF via ages and luminosities of the brightest clusters}
\label{sec:montecarlo}

\subsection{Uniform age distribution and a truncated power-law mass function}
\label{subsec:uniform}

\begin{figure}
\centering
\includegraphics[width=85mm]{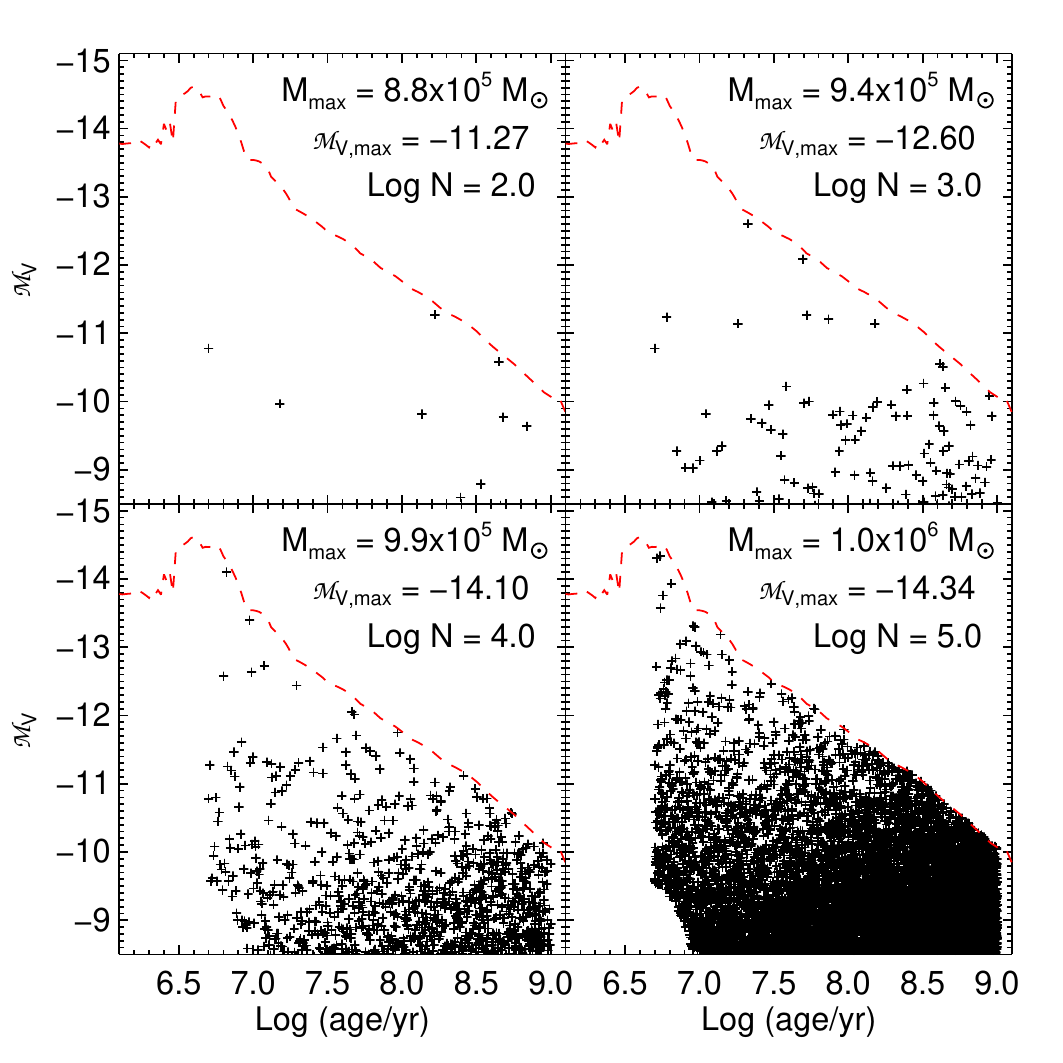}
\caption{\label{fig:logamvsim}Simulated absolute magnitude $\mathcal{M}_V$ versus log(age) plots for cluster systems of increasing richness. A power-law mass function truncated at $10^6$ $M_\odot$ is assumed.}
\end{figure}

As a first illustration of the statistical behaviour of the upper mass- and luminosity boundaries of a cluster population, clusters with masses between $10^4$ \mbox{$M_{\odot}$}\ and $10^6$ \mbox{$M_{\odot}$}  were drawn at random from a power-law mass distribution, $dN/dM \propto M^{-2}$.  Each cluster was assigned a random age between 5 Myr and 1 Gyr. Luminosities corresponding to the age and mass of each cluster were calculated using solar-metallicity simple stellar population (SSP) models from \citet[][hereafter BC03]{bc03}. These models will be used throughout this paper and assume a Salpeter-like ($dN/dM \propto M^{-2.35}$) stellar initial mass function (IMF) for $0.1 < M/\mbox{$M_{\odot}$} < 100$, which probably overestimates the number of low-mass stars. For more realistic \citet{kroupa02} or \citet{chab03} IMFs, the mass-to-light ratios (and corresponding masses quoted in this paper) would be smaller by a factor of about 1.5. 

Figure~\ref{fig:logamvsim} shows the resulting (log(age), $\mathcal{M}_V$) plots for samples of $N=10^2$, $10^3$, $10^4$ and $10^5$ clusters (note the use of calligraphic letters $\mathcal{M}_V$ in this paper to distinguish absolute magnitudes from masses). The dashed line in each panel is the fading line of a $10^6\mbox{$M_{\odot}$}$ cluster, the maximum possible brightness at a given age for the truncated power-law MF assumed here. For a Schechter function, there would be no strict upper limit. While the effects of cluster disruption are ignored in this plot, it serves to demonstrate a couple of important points. First, although the \emph{mass} function is sampled up to near the cut-off mass in all four cases, this is not the case for the \emph{luminosity} function. The brightest $\mathcal{M}_V$ magnitude possible for a $10^6$ $M_\odot$ cluster is $\mathcal{M}_V^{\rm max} = -14.5$, but only the two most populous samples get close to this value. In other words, \emph{the maximum cluster mass in each sample is close to the physical limit set by the MF, but the maximum  brightness is determined by random sampling}. This holds true over several orders of magnitude in $N$.  Second, the brightest cluster tends to be younger in the more populous cluster systems.  For a poor cluster system, rapid fading reduces the probability that a very massive cluster happens to be caught in the short time interval after its birth when it is most luminous. As the number of clusters is increased, it becomes increasingly likely to encounter a young cluster with a mass near the physical upper limit and a luminosity that cannot be matched by an older object. Thus, the mean or median age of the brightest cluster shifts towards younger values.

The specific age-luminosity relation for the brightest cluster will depend on the actual upper mass limit, whether this is a strict cut-off as in Fig.~\ref{fig:logamvsim} or the $\mbox{$M_c$}$ of a Schechter function. It therefore potentially offers a  diagnostic of the upper limit. This will be discussed in further detail below (Sect.~\ref{sec:details}). 

\subsection{Cluster disruption}
\label{sec:disrupt}

The simple experiment described in the previous paragraph suffers from a number of shortcomings. In particular, the relation between the age and magnitude of the brightest cluster will depend on the age distribution of the cluster population. This will, in general, be affected by variations in the star (cluster) formation rate as well as by disruption.

A typical open cluster in the Milky Way will gradually evaporate and dissolve completely within a few times $10^8$ years \citep{wielen71}, due to the combined effects of stellar evolution, internal two-body relaxation and external perturbations \citep{spit58,spit87,terle87,lg08}.  This ``secular evolution'' \citep{pf08} causes a cluster to lose mass at a (roughly) constant rate, so that lifetimes are in general expected to scale with cluster mass. The exact relation between lifetime and mass will also depend on the structural parameters of the clusters, the relation between half-mass radius and mass, and the environment in which they find themselves. According to \citetalias{lam05}, a useful fitting formula is
\begin{equation}
  t_{\rm dis} = t_4 \left(\frac{M_i}{10^4 \mbox{$M_{\odot}$}}\right)^\gamma
  \label{eq:lamers}
\end{equation}
where the parameter $\gamma$ has a value of about 0.62 and $t_{\rm dis}$ is the disruption time scale of a cluster with initial mass $M_i$. For the Milky Way, \citetalias{lam05} find a disruption time scale for a $10^4\mbox{$M_{\odot}$}$ cluster of $t_4 = 1.3\times10^9$ years, so that extrapolation of Eq.~(\ref{eq:lamers}) leads to a lifetime of $\sim5$ Gyr for a $10^5 \mbox{$M_{\odot}$}$ cluster and $\sim23$ Gyr for a $10^6 \mbox{$M_{\odot}$}$ cluster. Thus, the most massive clusters are expected to survive for a significant fraction of the age of the Universe in the Galactic disc. It should be noted that a $\gamma$-value closer to unity is required to reproduce the MF of  old GCs as a result of dynamical evolution from an initial power-law or Schechter function \citep[e.g.][]{jordan07}. It therefore remains uncertain whether a universal $\gamma$ value applies everywhere, but for our purpose the exact numerical value is not important.

Many clusters are likely born out of virial equilibrium, or are forced into a state of non-equilibrium once residual gas from the star formation process is expelled. During the ensuing violent relaxation, clusters may disrupt completely or partially and undergo a phase of ``infant mortality'' or ``infant weight loss'' which lasts up to a few tens of Myr \citep{kb02,ll03,gb06,whit07}. Unlike the secular dissolution discussed above, this process is believed to affect clusters of all masses more or less equally. In the Milky Way, the Magellanic Clouds, and possibly several other galaxies, mass-independent disruption seems to be relatively unimportant for non-embedded clusters older than $\sim10$ Myr (\citealt{lam05,gieles07b,deg08,gibast08}, but see \citealt{cfw06} for an alternative view regarding the SMC). For the Antennae galaxies, mass-independent disruption instead seems to lead to the loss of 80--90\% of the cluster population per logarithmic age bin for the first $\sim10^8 -10^9$ years \citep{fcw05,whit07}.

For secular dissolution the conclusions drawn from Fig.~\ref{fig:logamvsim} would not change dramatically unless the disruption time scale is much shorter than in the Milky Way. Even for a disruption time as short as $t_4=200$ Myr, as derived for the inner disc of NGC~5194 \citep{gieles05}, a $10^5 \mbox{$M_{\odot}$}$ cluster would still survive for about 800 Myr (although its mass would be steadily decreasing over this time span). Mass-independent disruption, on the other hand, can have a strong impact on the age distribution of any subsample. In the following, both mass-dependent and mass-independent disruption will be considered.

\subsection{Including cluster disruption and Schechter mass functions: the set-up}
\label{sec:details}

\begin{figure}
\centering
\includegraphics[width=85mm]{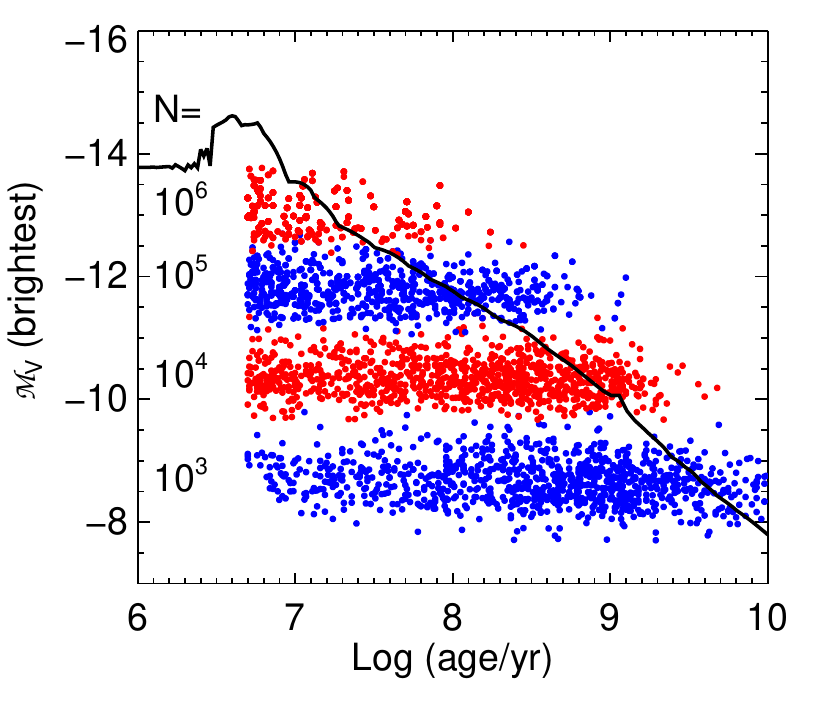}
\caption{\label{fig:scat}Absolute $\mathcal{M}_V$ magnitude versus age for the brightest cluster when sampling $N$ clusters at random from a Schechter mass function with $\mbox{$M_c$} = 10^6$ $M_\odot$. For each $N$, a number of realisations are shown. The curve is the $\mathcal{M}_V$ magnitude vs.\ age for a $10^6\mbox{$M_{\odot}$}$ cluster. No disruption is included.}
\end{figure}

Fig.~\ref{fig:scat} shows the median $\mathcal{M}_V$ vs.\ log(age) for the brightest cluster in random samples of $N=10^3, 10^4, 10^5$ and $10^6$ clusters with $M>5000\mbox{$M_{\odot}$}$ drawn from a Schechter mass function with $\mbox{$M_c$}=10^6\mbox{$M_{\odot}$}$ and $\alpha=-2$. In this case, no cluster disruption was included and the age distribution was assumed to be uniform over $10^{10}$ years. As expected, $\mathcal{M}_V$ generally becomes brighter, and log(age) younger, when sampling more clusters. It is perhaps not quite intuitive that the $\mathcal{M}_V$ and age of the brightest cluster should be uncorrelated for fixed $\mbox{$M_c$}$ and $N$. This is due to the fact that clusters are selected based on their luminosities, so that the scatter in log(age) in Fig.~\ref{fig:scat} simply represents the range of possible ages for a given $\mathcal{M}_V$. For small $N$ the age distribution is very extended and the brightest cluster can be either young (and relatively low mass) or old and relatively massive. For large $N$ the brightest cluster will generally be fairly young and massive. Also included in the figure is the \citetalias{bc03} model prediction for the $\mathcal{M}_V$ magnitude of a $10^6\mbox{$M_{\odot}$}$ cluster as a function of age. Since the Schechter function does not cut off abruptly at \mbox{$M_c$}, the brightest cluster will be more massive than \mbox{$M_c$} in some fraction of the realisations. This occurs more frequently in rich cluster systems where the high-mass tail of the Schechter function is more fully sampled over the entire age range, but can happen occasionally even in a cluster-poor galaxy. For the Monte-Carlo simulations shown in Fig.~\ref{fig:scat}, the brightest cluster is more massive than $\mbox{$M_c$}$ in 39\% of the realisations for $N=10^6$, and in 7\% of the cases for $N=10^3$. This again illustrates that while the luminosity of the brightest cluster is strongly dependent on $N$, this does not translate trivially to a similar scaling of the maximum cluster mass with $N$.

To quantify the effects of cluster disruption, three cases are considered:
\begin{itemize}
  \item Case A: No disruption
  \item Case B: No infant mortality, and secular evolution with $t_4 = 10^8$ years and exponent $\gamma = 0.62$. The disruption time scale assumed here is deliberately chosen to be much shorter (a factor of 10) than derived for the Solar neighbourhood, so as to illustrate the effect of secular disruption in an extreme case. In this scenario, the mass $M(t)$ at time $t$ of a cluster with initial mass $M_i$ can be approximated by
   \begin{equation}
     M(t) \simeq M_i \left[ (\mu_{\rm ev}(t))^\gamma - \frac{\gamma t}{t_4} \left( \frac{10^4 M_\odot}{M_i}\right)^\gamma\right]^{1/\gamma}
     \label{eq:secdis}
   \end{equation}
   where $\mu_{\rm ev}(t)$ indicates the fraction of mass left in the cluster after accounting for mass loss due to stellar evolution.
 This quantity is tabulated in the \citetalias{bc03} models. Eq.~(\ref{eq:secdis}) can be inverted to 
   \begin{equation}
     M_i = 10^4 \mbox{$M_{\odot}$} \left[\left(\frac{M(t)}{10^4\mbox{$M_{\odot}$}}\right)^\gamma + \frac{\gamma t}{t_4}\right]^{1/\gamma} \mu_{\rm ev}(t)^{-1}
     \label{eq:secdisinv}
   \end{equation}

  \item Case C: Infant mortality leading to the loss of IMR=90\% of clusters per log(age) interval until $\tau_{\rm im,max} = 10^8$ years and no secular evolution. For a  mass-limited sample, the surviving fraction of clusters $f_{\rm surv}$ will be
  \begin{equation}
    f_{\rm surv} = \left\{ \begin{array}{ll}
     1 & \mbox{for $\tau \leq \tau_0$} \\
      (\tau/\tau_0)^{\log (1 - {\rm IMR})} & \mbox{for $\tau_0 < \tau < \tau_{\rm im,max}$} \\
      (\tau_{\rm im,max}/\tau_0)^{\log(1 - {\rm IMR})} & \mbox{for $\tau \ge \tau_{\rm im,max}$}
    \end{array}
    \right.  
  \end{equation}
where $\tau_0$ marks the onset of infant mortality. We assume $\tau_0 = 5\times10^6$ yr so $f_{\rm surv} = \left[\tau/(5\times10^6 \, {\rm years})\right]^{-1}$ for the first $10^8$ years and $f_{\rm surv} = 0.05$ afterwards \citep{whit07}.
\end{itemize}

\begin{figure}
\centering
\includegraphics[width=85mm]{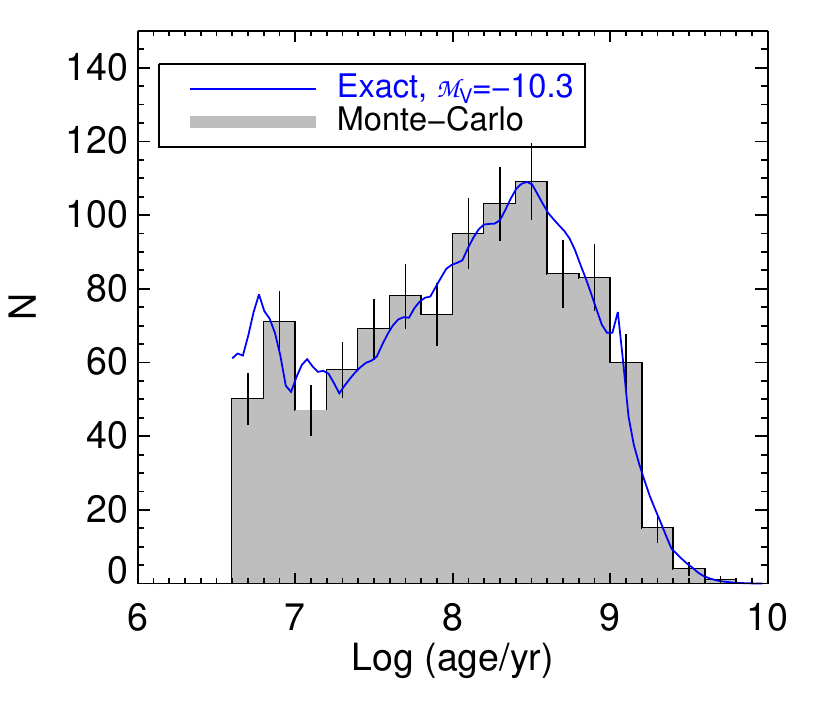}
\caption{\label{fig:scathist}Age distribution of the brightest cluster for $\mbox{$M_c$}=10^6\mbox{$M_{\odot}$}$ and $N=10^4$. The histogram is for the $N=10^4$ Monte-Carlo simulation in Fig.~\ref{fig:scat} while the solid (blue) curve shows the exact relation according to Eq.~(\ref{eq:dens}) for $\mathcal{M}_V=-10.3$ mag.}
\end{figure}

Monte-Carlo simulations similar to those shown in Fig.~\ref{fig:scat} could, in principle, be used to determine how $\mathcal{M}_V$ and log(age) for the brightest cluster depend on \mbox{$M_c$} and $N$ for each disruption scenario. However, this can also be established more generally by considering the distribution of cluster ages corresponding to a particular absolute magnitude, $\mathcal{M}_V$, or luminosity $L_V$. For  a sample of clusters with fixed age $\tau$, the luminosity function will be
\begin{equation}
   \left. \frac{dN}{dL_V}\right|_\tau = \frac{dN}{dM} \left. \frac{dM}{dL_V} \right|_\tau
   \label{eq:dndl}
\end{equation}
where here $M(L_V,\tau)$ is the mass of a cluster with age $\tau$ and ($V$-band) luminosity $L_V$. The first factor expands to
\begin{equation}
  \frac{dN}{dM} = \frac{dN}{dM_i} \frac{dM_i}{dM}
\end{equation}
where $dN/dM_i$ is the ICMF and $dM_i/dM$ follows by straight forward differentiation of Eq.~(\ref{eq:secdisinv}). The second factor in Eq.~(\ref{eq:dndl}) is the mass-to-light ratio, $\Upsilon_V(\tau) \equiv M(L_V,\tau)/L_V$. Representing the ICMF by $\psi(M_i)$, we can then write 
%
%
%
\begin{equation}
  \frac{d^2N}{d\tau\,dL_V} =  \psi(M_i) \left(\frac{dM_i}{dM} \right) \Upsilon_V(\tau) \, {\rm CFR}(\tau) \, f_{\rm surv}(\tau)
  \label{eq:dens}
\end{equation}
where CFR is the cluster formation rate (of bound clusters) and $f_{\rm surv}(\tau)$ is the survival fraction after applying mass-independent disruption. It is assumed here that $\Upsilon$ only depends on $\tau$ so that it can be computed from standard SSP models, but a mass dependency \citep[e.g.,][]{kruij08} could in principle also be included. In general, the ICMF should be normalised to unit mass per unit time over some range of initial cluster masses $M_{\rm lo} < M_i < M_{\rm up}$.
%
%

Fig.~\ref{fig:scathist} compares the age distribution for the Monte-Carlo simulation for $\mbox{$M_c$}=10^6 M_\odot$ and $N=10^4$ (Fig.~\ref{fig:scat}) with the exact expression (\ref{eq:dens}), scaled to match the Monte-Carlo data. We have used $\mathcal{M}_V=-10.3$, which is the median absolute magnitude of the brightest cluster for this $\mbox{$M_c$}$ and $N$.  The Monte-Carlo simulations and exact expression show excellent agreement within the Poisson errors.  From Eq.~(\ref{eq:dens}) it is straight forward to calculate statistics such as the mean or median age of clusters with a given luminosity, as well as the standard deviation, mean absolute deviation, etc. These statistics will apply to \emph{any} cluster sample selected by luminosity, i.e.\ not only the brightest cluster, but also the \mbox{$5^{\rm th}$} brightest, or clusters with a fixed magnitude. In addition, the integral over all $\tau$ gives the luminosity function, $dN/dL_V$. 

\subsection{Predicted median (log(age), $\mathcal{M}_V$) relations for the different disruption scenarios}

\begin{figure*}
\centering
\includegraphics[height=180mm,angle=90]{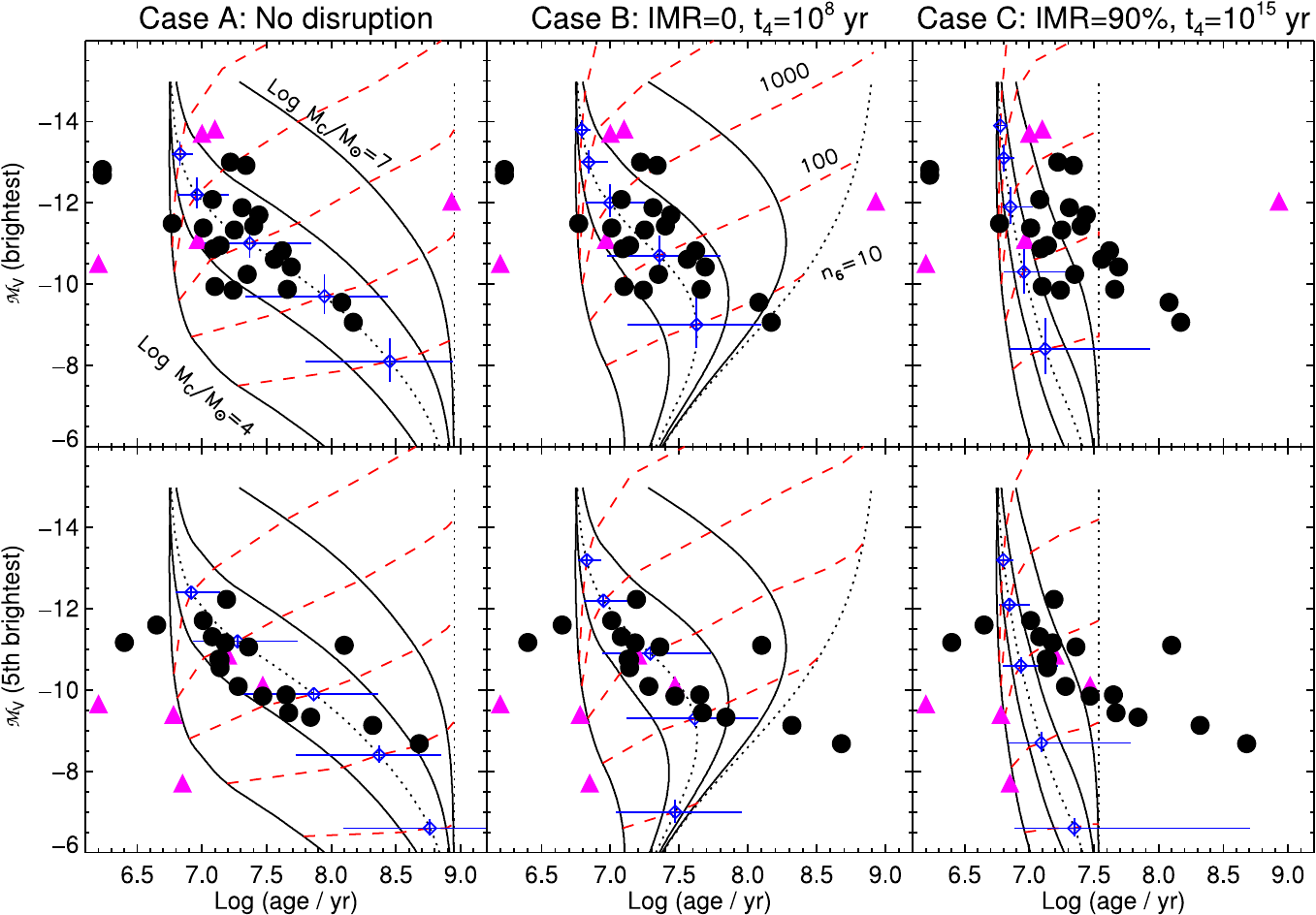}
\caption{\label{fig:logamvobs_all}The median absolute magnitude of the brightest (top) and \mbox{$5^{\rm th}$} brightest (bottom) cluster versus median age.  Results are shown for Schechter mass functions with \mbox{$M_c$} = $10^4$, $10^5$, $10^6$ and $10^7$ $M_\odot$ and $\alpha=-2$ (solid curves), as well as for $\mbox{$M_c$}=3\times10^5 \mbox{$M_{\odot}$}$ and $\mbox{$M_c$}=10^{15}\mbox{$M_{\odot}$}$ (dotted curves). The (red) dashed lines are for constant total number of clusters ($n_6=10^1$, $10^2$, \ldots, $10^5$) brighter than $\mathcal{M}_V=-6$. The error bars mark the median absolute deviation of log(age) and $\mathcal{M}_V$ for fixed $\mbox{$M_c$}$ and $n_6$. The triangles mark clusters in irregular galaxies while circles are for spirals (see Table~\ref{tab:data}).
}
\end{figure*}

The median (log(age), $\mathcal{M}_V$) relations for the three disruption scenarios are shown in Fig.~\ref{fig:logamvobs_all}. The relations have been calculated for Schechter functions with $\alpha=-2$ and $\mbox{$M_c$}=10^4\mbox{$M_{\odot}$}, 10^5\mbox{$M_{\odot}$}, 10^6\mbox{$M_{\odot}$}$ and $10^7\mbox{$M_{\odot}$}$ (solid lines) as well as for $\mbox{$M_c$}=3\times10^5\mbox{$M_{\odot}$}$ and $\mbox{$M_c$}=10^{15} \mbox{$M_{\odot}$}$ (dotted lines). The latter  is essentially an untruncated power-law with $\alpha=-2$. An age range of 5 Myr -- 10 Gyr and a constant CFR are assumed in all cases. In addition to the constant-$\mbox{$M_c$}$ curves, lines for constant \emph{number} of clusters ($n_6$) brighter than $\mathcal{M}_V=-6$ are drawn with (red) dashed lines. Note the use of lower-case $n$ to distinguish this from the number of clusters above a particular \emph{mass} limit, $N$, which is more difficult to relate directly to observations. The constant-$n_6$ curves were determined by first finding a normalisation constant $n_0$ for the luminosity functions so that
\begin{equation}
  n_6 = n_0 \, \int^{\mathcal{M}_V=-6}_{\mathcal{M}_V=-\infty} \frac{dN}{d\mathcal{M}_V} \, d\mathcal{M}_V
\end{equation}
and then solving
\begin{equation}
  n_{\mathcal{M}_V < \rm br}  = n_0 \int_{\mathcal{M}_V=-\infty}^{\mathcal{M}_{V,n_6}^{\rm br}} \frac{dN}{d\mathcal{M}_V} \, d\mathcal{M}_V
  \label{eq:brightest}
\end{equation}
for the magnitude of the brightest cluster, $\mathcal{M}_{V,n_6}^{\rm br}$ and
\begin{equation}
  n_{\mathcal{M}_V < \rm 5th} = n_0 \int_{\mathcal{M}_V=-\infty}^{\mathcal{M}_{V,n_6}^{\rm 5th}} \frac{dN}{d\mathcal{M}_V} \, d\mathcal{M}_V
  \label{eq:fifth}
\end{equation}
for $\mathcal{M}_{V,n_6}^{\rm 5th}$, the magnitude of the \mbox{$5^{\rm th}$} brightest cluster. The constants on the left-hand sides of Eq.~(\ref{eq:brightest}) and Eq.~(\ref{eq:fifth}) were found by comparison with Monte-Carlo experiments where $n_6$ clusters were sampled at random from the LFs. Based on 1000 such samples for each of a variety of disruption scenarios, $\mbox{$M_c$}$ and $n_6$ values, $n_{\mathcal{M}_V<\rm br}=0.75$ and $n_{\mathcal{M}_V<\rm 5th}=5$ were found to be suitable values. With these constants, Eq.~(\ref{eq:brightest}) and (\ref{eq:fifth}) reproduced the median magnitudes of the brightest and \mbox{$5^{\rm th}$} brightest clusters within better than 0.1 mag. The Monte-Carlo experiments were also used to estimate the vertical error bars in Fig.~\ref{fig:logamvobs_all}, which indicate the median absolute deviation of the brightest and \mbox{$5^{\rm th}$} brightest cluster magnitudes for fixed $n_6$ and $\mbox{$M_c$}$. Note that the scatter in $\mathcal{M}_V$ is significantly smaller for the \mbox{$5^{\rm th}$} brightest cluster. The horizontal error bars, indicating the median absolute deviation of $\tau$, were calculated directly by integrating Eq.~(\ref{eq:dens}) over $\tau$ to find the positive and negative age intervals with respect to the median that include half the clusters for a given $\mathcal{M}_V$. The sizes of the horizontal error bars are given simply by the shape of the age distribution, and are therefore identical for the brightest and \mbox{$5^{\rm th}$} brightest clusters. Because the constant-$\mbox{$M_c$}$ curves are fixed by the age distribution of clusters for a given magnitude, the model relations in the upper and lower panels differ only in the constant-$n_6$ (dashed) curves. These are shifted to fainter magnitudes in the lower panels, in accordance with the median magnitude difference between the brightest and \mbox{$5^{\rm th}$} brightest cluster for each ($\mbox{$M_c$}, n_6$) combination.

In the special case where there is no disruption and the MF is a uniform power-law, the age distribution is luminosity-independent. In this case, the median age of the brightest cluster (or any other luminosity-selected sample) will also be independent of luminosity and $n_6$, as indicated by the vertical dashed line for the Case A scenario in Fig.~\ref{fig:logamvobs_all}. This is true regardlessly of the actual value of the MF slope $\alpha$, although the specific median age will depend on $\alpha$. For Schechter MFs with a finite $\mbox{$M_c$}$, the median age will instead decrease as $n_6$ and the luminosity of the brightest cluster increase, as discussed above and illustrated by the solid lines in the Case A panels. The introduction of secular dissolution does not strongly affect the age distribution of the brightest clusters in cluster-rich systems, where these clusters will tend to be massive and long-lived. Hence, the Case A and Case B scenarios resemble each other for large $n_6$. However, for very cluster-poor systems secular dissolution starts to have a noticeable effect on the age distribution of even the brightest clusters, and the median age shifts back towards younger ages as $n_6$ becomes small. For the Case C scenario the age distribution is, per definition, deficient in older clusters of all masses. In this case the median age shifts to younger values for all $n_6$, albeit with a tail towards older ages. 

For Case A and Case B, Fig.~\ref{fig:logamvobs_all} shows a clear separation between the curves for different \mbox{$M_c$}, suggesting that this type of diagram may indeed be used as a diagnostic tool to constrain \mbox{$M_c$}.  However, it is equally clear that much of this diagnostic power is lost for cluster samples that are  dominated by a high degree of mass independent disruption, as in Case C.  Due to the large scatter, one should not assign much significance to the location of any particular galaxy  in  Fig.~\ref{fig:logamvobs_all}, but for large galaxy samples it is still possible to draw conclusions about \mbox{$M_c$} in a statistical sense. The inconvenience of having to include a large number of galaxies is counterbalanced by the fact that only the brightest (or few brightest) clusters have to be identified and age-dated. Furthermore, it is not a strict requirement that the entire galaxy is covered; any sub-sample of clusters can equally well be compared with the grids, as long as the sub-sample can be assumed to be representative of the global cluster population in a galaxy. Of course, this assumption may not always be justified.

It is important to stress that the grids in Fig.~\ref{fig:logamvobs_all} are only suitable for comparison with galaxies for which it is plausible that the cluster formation rate has been approximately constant over the past few Gyrs. More specifically, the cluster population should not be dominated by a single strong burst so galaxies which are currently undergoing strong bursts of star formation, or have done so in the past (e.g.\ merger remnants) would require a custom-made grid tailored for their specific star formation histories. However,  variations in the cluster formation rate of a factor of a few are still acceptable. Among normal disc galaxies, which are our primary interest here, star- and cluster formation histories are more likely to have been roughly uniform, at least on average. For the  Milky Way, several studies suggest that variations of a factor of 2--3 in the star formation rate on time scales of $10^8$ - $10^9$ years \citep{rocha00,her00} are applicable.

\subsection{Comparison with observations}
\label{subsec:data}

\begin{table}
\begin{minipage}[t]{\columnwidth}
\caption{Data for the brightest and \mbox{$5^{\rm th}$} brightest clusters in a sample of galaxies.
}
\label{tab:data}
\renewcommand{\footnoterule}{}  
\begin{tabular}{lcccccc} \hline \hline
Galaxy   & \multicolumn{2}{c}{Brightest} & \multicolumn{2}{c}{\mbox{$5^{\rm th}$} brightest} & Ref \footnote{References:
1: \citet{and04},
2: \citet{bhe02},
3: \citet{gel01},
4: \citet{lar99},
5: \citet{lar02}} & N\footnote{Number of clusters detected} \\
          &   $\mathcal{M}_V$  & log(age)   & $\mathcal{M}_V$  & log(age) & \\ \hline
Irregulars: \\
\object{NGC 1569}  & $-13.82$ & 7.10 &  $-10.86$  &  7.20  & 1 & - \\
\object{NGC 1705}  & $-13.71$ & 7.00 &  $-7.71 $  &  6.85  & 2 & - \\
\object{NGC 4214}  & $-12.04$ & 8.93 &  $-9.40 $  &  6.78  & 2 & - \\
\object{NGC 4449}  & $-10.51$ & 6.2  &  $-9.66 $  &  6.2   & 3 & - \\
\object{NGC 1156}  & $-11.10$ & 6.97 &  $-10.12$  &  7.47  & 4 & 22 \\
Spirals: \\
\object{NGC 45}    & $-9.94 $ & 7.10 &     -      &   -    & 4 & 3 \\
\object{NGC 247}   & $-10.24$ & 7.35 &     -      &   -    & 4 & 3 \\
\object{NGC 300}   & $-9.85 $ & 7.24 &     -      &   -    & 4 & 3 \\
\object{NGC 628}   & $-11.33$ & 7.25 &  $-10.55$  &  7.14  & 4 & 38 \\
\object{NGC 1313}  & $-12.08$ & 7.08 &  $-11.16$  &  7.18  & 4 & 45 \\
\object{NGC 2403}  & $-9.87 $ & 7.66 &  $ -9.44$  &  7.67  & 4 & 14 \\
\object{NGC 2835}  & $-10.87$ & 7.09 &  $ -9.88$  &  7.65  & 4 & 9 \\
\object{NGC 2997}  & $-12.92$ & 7.34 &  $-12.23$  &  7.19  & 4 & 34 \\
\object{NGC 3184}  & $-10.61$ & 7.56 &  $ -9.13$  &  8.32  & 4 & 13 \\
\object{NGC 3521}  & $-11.49$ & 6.77 &  $-11.10$  &  8.10  & 5 & 13 \\
\object{NGC 3621}  & $-11.88$ & 7.31 &  $-11.06$  &  7.36  & 4 & 45 \\
\object{NGC 4258}  & $-12.68$ & 6.23 &  $-11.17$  &  6.40  & 5 & 44 \\
\object{NGC 4395}  & $-9.06 $ & 8.17 &     -      &   -    & 4 & 2 \\
\object{NGC 5055}  & $-11.38$ & 7.01 &  $-10.75$  &  7.14  & 5 & 24 \\
\object{NGC 5194}  & $-12.82$ & 6.23 &  $-11.60$  &  6.65  & 5 & 69 \\
\object{NGC 5204}  & $-9.55 $ & 8.08 &  $ -8.68$  &  8.68  & 4 & 7 \\
\object{NGC 5236}  & $-11.70$ & 7.44 &  $-11.31$  &  7.08  & 4 & 149 \\
\object{NGC 5585}  & $-10.82$ & 7.62 &  $ -9.33$  &  7.84  & 4 & 9 \\
\object{NGC 6744}  & $-10.95$ & 7.14 &  $-10.09$  &  7.28  & 4 & 18 \\
\object{NGC 6946}  & $-13.00$ & 7.22 &  $-11.71$  &  7.01  & 4 & 107\\
\object{NGC 7424}  & $-11.43$ & 7.40 &  $-10.76$  &  7.13  & 4 & 10 \\
\object{NGC 7793}  & $-10.42$ & 7.69 &  $ -9.85$  &  7.47  & 4 & 20 \\
\hline
\end{tabular}
\end{minipage}
\end{table}

Table~\ref{tab:data} lists data for the brightest and \mbox{$5^{\rm th}$} brightest cluster in a sample of galaxies. These are mostly spirals, taken from our previous work, with the addition of a few irregular galaxies.  For the galaxies with data from \citet{lar99,lar02}, which are ground-based with more or less homogeneous detection limits, the last column lists the total number of clusters detected in each galaxy. Formally, the detection limit is at $\mathcal{M}_V=-8.5$ ($\mathcal{M}_V=-9.5$ for clusters bluer than $U\!-\!B = -0.4$), but significant incompleteness likely sets in well above these limits. In selecting the cluster samples in the spirals, objects with H$\alpha$ emission were excluded so that the samples mostly consist of clusters older than about 10 Myr. 

Overall properties of the cluster systems and the host galaxies have been discussed extensively by \citet{lr00}. The spiral galaxies span more than an order of magnitude in absolute and area-normalised star formation rate, and are mostly type Sbc or later. With the exception of NGC~5194, none is involved in any strong on-going interaction. The star formation rates show a strong correlation with the richness of the cluster systems \citep{lar02}, so that the last column in Table~\ref{tab:data} may also be taken as an indication of the relative star formation rates in the galaxies. The irregular galaxies constitute a more heterogeneous sample, with at least some showing clear evidence of bursty star formation histories.

A requirement in collecting data for Table~\ref{tab:data} was that age estimates and $V$-magnitudes were available for each individual cluster. In a few cases where less than five clusters were detected, the columns for the \mbox{$5^{\rm th}$} brightest cluster are left empty.  Ages were obtained from $UBV$ photometry, employing the S-sequence method \citep{gir95}. The S-sequence is essentially a curve drawn through the mean colours of Large and Small Magellanic Cloud clusters in the ($U\!-\!B$, $B\!-\!V$) plane. Age increases along the curve from blue to red colours, and clusters which do not fall exactly on the curve are projected onto it along straight lines that take into account the effects of stochastic colour fluctuations as well as interstellar extinction. Alternatively, ages could have been estimated by comparing the observed colours with SSP models. This was also tried (using the \citetalias{bc03} models), and differences with respect to the S-sequence were found to be minor. The latter was preferred here due to its empirical foundation, although the S-sequence is strictly valid only for LMC-like metallicity. For further details concerning the properties of individual galaxies and the data reduction and analysis, please consult the references given in Table~\ref{tab:data}.

The spirals are overplotted with filled circles in Fig.~\ref{fig:logamvobs_all} while the irregulars are shown with triangles.  Comparing with the Case A and B panels, most data points fall between the $\mbox{$M_c$}=10^5\mbox{$M_{\odot}$}$ and $\mbox{$M_c$}=10^6\mbox{$M_{\odot}$}$ curves. The two most outlying data points are for NGC~1569 and NGC~1705, both of which have experienced recent bursts of star formation \citep{anni03,ange05} and are dominated by one or two very bright, young clusters. These two systems therefore provide a good illustration of the limitations of this method. For the rest of the data points, a Schechter function with an $\mbox{$M_c$}$ of a few times $10^5\mbox{$M_{\odot}$}$ fits the mean trend quite well for both the brightest and \mbox{$5^{\rm th}$}-brightest cluster. An exception to this statement occurs at low $n_6$ in the Case B scenario, which may well be an indication that the disruption time scale assumed here is too short to match these galaxies. For the Case C scenario the picture is less clear. No single $\mbox{$M_c$}$ value can match the data for the disruption parameters assumed here. A lower IMR would alleviate this problem to some extent by making the (log(age), $\mathcal{M}_V$) relations more similar to those for the no-disruption scenario. For example, an IMR of 50\% per dex over $10^9$ years would provide an acceptable fit. Alternatively, the trend towards higher median ages in low-$n_6$ galaxies might be explained by invoking a lower infant mortality rate in these galaxies specifically, causing a shallower decline in $f_{\rm surv}$ with age.

Of the galaxies in Table~\ref{tab:data}, cluster disruption has only been studied in a few. In NGC~5194, the age-mass distribution of clusters older than about 10 Myr can be well fitted by a secular evolution 
model with a rather short disruption time scale of $t_4\approx 200$ Myr  \citep{bast05,gieles05}. Five additional galaxies (NGC~45, NGC~1313, NGC~4395 NGC~5236, NGC~7793) have been studied by \citet{mora08}, but their data did not allow a clear distinction between secular evolution and infant mortality disruption scenarios.  Without detailed knowledge about the cluster disruption laws in the galaxies, the interpretation of Fig.~\ref{fig:logamvobs_all} in terms of \mbox{$M_c$} remains somewhat ambiguous. However, it does seem clear that if the MF is a Schechter function, then \mbox{$M_c$} cannot be much lower than a few times $10^5 \mbox{$M_{\odot}$}$ in any of these galaxies, regardless of the cluster disruption law. More generally, any significant steepening (or truncation) of the MF below a few times $10^5 \mbox{$M_{\odot}$}$, compared to a power-law with a slope around $-2$, is ruled out.

\subsection{Radial trends}

\begin{figure}
\centering
\includegraphics[width=85mm]{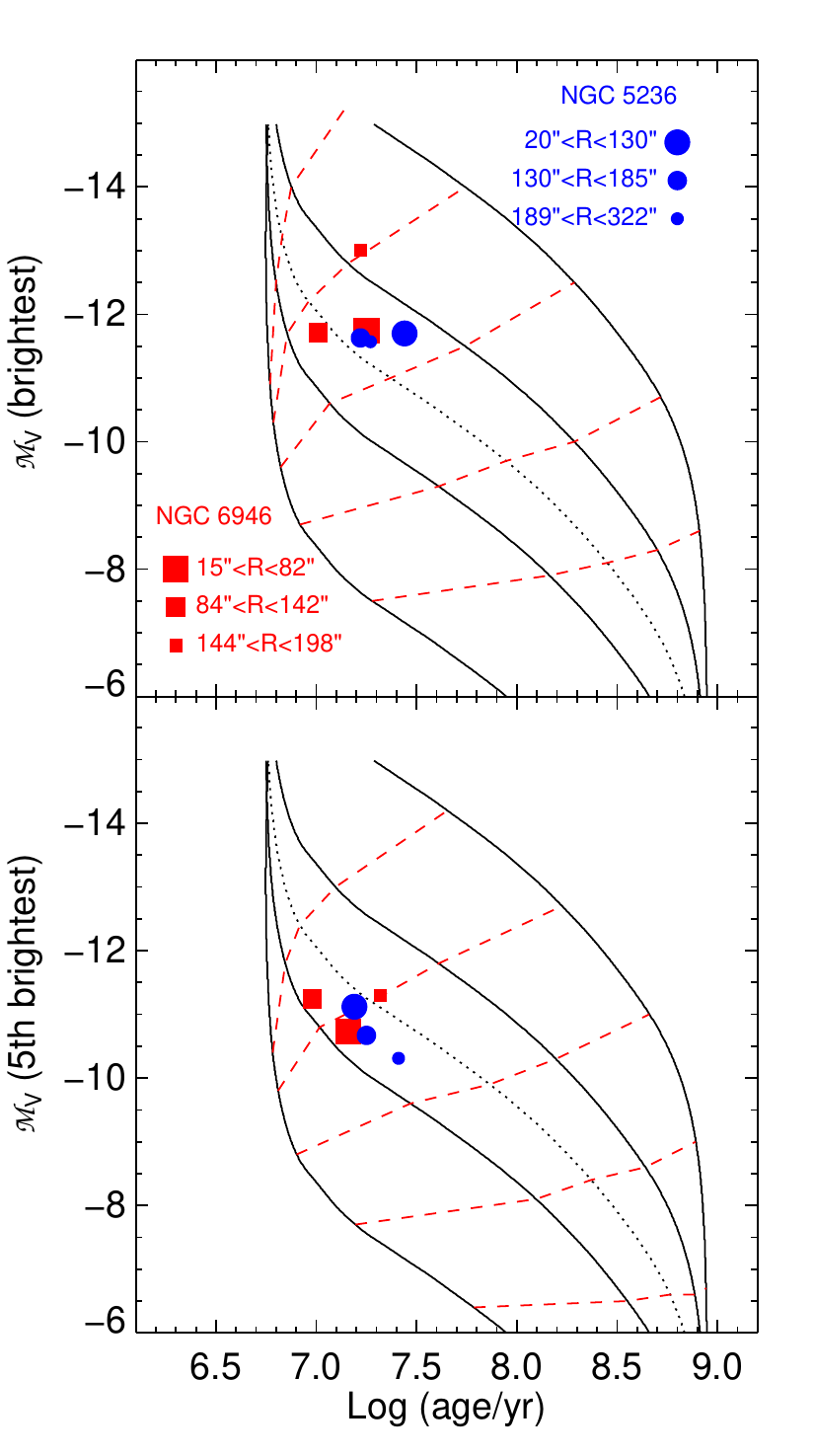}
\caption{\label{fig:subsamp_io}$\mathcal{M}_V$ versus log age for the brightest and \mbox{$5^{\rm th}$} brightest clusters in radial sub-samples in NGC~5236 and NGC~6946. In each galaxy, each radial bin contains an equal number of clusters. The model lines are the same as for the Case A scenario in Fig.~\ref{fig:logamvobs_all}. }
\end{figure}

Many properties of galaxies correlate with galactocentric distance: gas density, star formation rate, relative velocities of spiral arms and stellar orbits, metallicity, etc. It is not unreasonable to suspect
that this might also lead to radial variations in the cluster MF.  So far, radial trends in the properties of (young) cluster populations remain poorly studied. In NGC~5194, a radially dependent bend in the LF suggests that the upper limit of the MF might decrease outwards \citep{haas08}. 

Fig.~\ref{fig:subsamp_io} shows the same Case A model grids as the left-hand column in Fig.~\ref{fig:logamvobs_all}, together with the age and magnitude of the brightest and \mbox{$5^{\rm th}$} brightest cluster
in three radial subsamples for NGC~5236 and NGC~6946. These two galaxies have the largest number of detected clusters among those in Table~\ref{tab:data}. The radial bins were defined such that each bin contains an equal number of clusters (49 in NGC~5236, 35 in NGC~6946). The corresponding radial ranges are indicated in the figure legend. As in Fig.~\ref{fig:logamvobs_all}, the scatter of the data points is well within the error bars, i.e.\ the data are consistent with a radially independent \mbox{$M_c$} in both NGC~5236 and NGC~6946.

\section{Comparing MFs directly}
\label{sec:mfs}

\begin{figure}
\centering
\includegraphics[width=85mm]{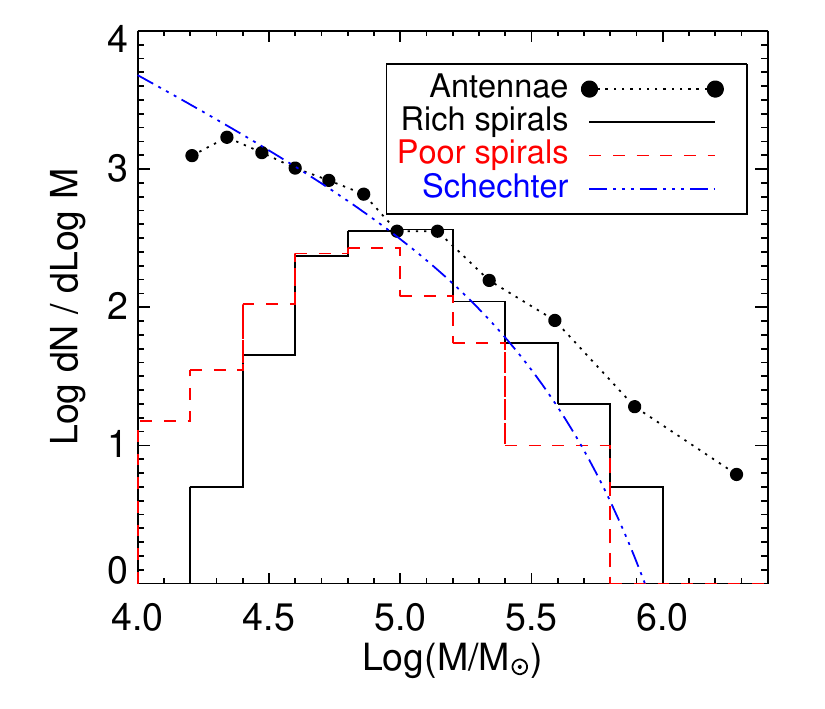}
\caption{\label{fig:mf_spir}Combined cluster mass functions for the two galaxies with more than 100 clusters each (NGC~5236, NGC~6946) and those with less than 40 clusters each. Clusters with ages $< 2\times10^8$ years are included. Also shown is the mass function for young clusters in the Antennae \citep{zf99} and a Schechter function with $\mbox{$M_c$} = 2.1\times 10^5 \mbox{$M_{\odot}$}$ and $\alpha=-2$.
}
\end{figure}

The ground-based data used in Sect.~\ref{sec:montecarlo} are, unfortunately, not ideal for direct studies of the MF. Visual inspection played a central role in the identification of cluster candidates, so completeness effects are hard to quantify. Nevertheless, it is tempting to attempt a comparison of MFs derived from these data with a Schechter function.

First, a purely differential comparison was carried out by dividing the spiral galaxies in Table~\ref{tab:data} into a cluster-rich and a cluster-poor sample. The cluster-rich sample consists of those galaxies with more than 100 clusters each (NGC~5236, NGC~6946) and the cluster-poor sample of galaxies with less than 40 clusters each. NGC~2997 formally falls in the cluster-poor category, but has been excluded because it is more distant than most of the other galaxies in the sample.  Its high specific $U$-band cluster luminosity, $T_L(U)=1.45$  \citep{lr00} suggests that it would probably have fallen in the cluster-rich sample if located at the distance of NGC~5236 or NGC~6946. 

Having an estimate of the age and $\mathcal{M}_V$ for each cluster, masses were derived using mass-to-light ratios from \citetalias{bc03}. In principle, both the S-sequence method and a comparison with SSP models allow the extinction to be estimated for each individual cluster, by determining the shift along the extinction vector that is required to obtain the best match between observed and model/S-sequence colours. Both methods were tried, and yielded small mean $A_V$ values for both the cluster-poor and cluster-rich samples  ($\langle A_V \rangle \approx -0.1- +0.2$ mag). However, the scatter was large (0.4--0.6 mag for the S-sequence method and 0.7--1.0 mag when using the SSP models) with a significant tail towards negative extinctions. This suggests that a correction for extinction for each individual cluster would introduce significant spurious scatter, possibly because differences between observed and model colours can have other causes (e.g., stochastic colour fluctuations due to the finite number of stars in a cluster, or simply photometric errors). A choice was therefore made to omit extinction corrections for individual clusters.  However, the following analysis was, in fact, also carried out (though not described in detail here) for extinction-corrected cluster samples, with no change to the conclusions.

The MFs for clusters younger than 200 Myr in the two spiral samples are compared in Fig.~\ref{fig:mf_spir}. This age range was chosen such that secular evolution can be expected to play a relatively minor role in most galaxies.  At 200 Myr, the detection limit corresponds to a limiting mass of about  $7\times10^4 \mbox{$M_{\odot}$}$, but as in any cluster sample limited by a magnitude cut, this mass limit is strongly age-dependent and younger clusters with masses down to $\sim10^4\mbox{$M_{\odot}$}$ are present in the samples. Also shown is the MF for young clusters in the Antennae in the age range 25 Myr -- 160 Myr \citep{zf99}, derived assuming a distance of 19.2 Mpc. Recently, \citet{sav08} claimed a smaller distance of 13.3 Mpc based on the red giant branch tip, but \citet{schw08} find a distance of 22 Mpc based on observations of the Type-Ia supernova 2007sr and point out a number of problems with a smaller distance. Hence, we have kept the original mass scale of Fall \& Zhang. The apparent peaks in the MFs for the two spiral samples are undoubtedly due to incompleteness. The relatively higher fraction of clusters with masses below $10^5\mbox{$M_{\odot}$}$ in the cluster-poor spirals is most likely due the easier identification of low-mass clusters in these galaxies, which tend to have lower star formation rates, lower surface brightness, and less crowding. Above $10^5\mbox{$M_{\odot}$}$, however, the two spiral MFs appear remarkably similar to each other, and a Kolmogorov-Smirnov test confirms this impression by returning a $P$-value of 0.75 when comparing the two samples. In other words, \emph{the MFs in the cluster-rich and cluster-poor spirals are statistically indistinguishable}. Also worth noting is that the spiral MFs extend to nearly as massive clusters as those in the Antennae, although with a steeper slope.

A maximum-likelihood fit of a Schechter function with fixed faint-end slope $\alpha=-2$ to the combined spiral data returns a best-fit $\mbox{$M_c$} = (2.1\pm0.4)\times10^5 \mbox{$M_{\odot}$}$, also indicated in Fig.~\ref{fig:mf_spir} (arbitrarily scaled). Comparing this fit with the data, a K-S test returns a $P$-value of 0.22 for the combined spiral sample and $P=0.17$ and $P=0.86$ for the cluster-rich and cluster-poor samples separately. In all cases the mass distributions are consistent with being drawn from a single Schechter function. Moreover, the fitted $\mbox{$M_c$}$ value appears consistent when comparing with the Case A and B scenarios in Fig.~\ref{fig:logamvobs_all} and Fig.~\ref{fig:subsamp_io}. The fit is not unique: for example, a Schechter function with a steeper  faint-end slope ($\alpha=-2.2$) and a higher $\mbox{$M_c$}=2.5\times10^5\mbox{$M_{\odot}$}$ provides an equally good fit. However, a Schechter function with $\alpha=-2$ and $\mbox{$M_c$}=10^6\mbox{$M_{\odot}$}$ is ruled out at high confidence level; in this case a K--S test returns a $P$ value of only $2\times10^{-6}$ and even higher \mbox{$M_c$} are ruled out at still higher confidence level. Similarly, \mbox{$M_c$} values below $10^5\mbox{$M_{\odot}$}$ are also ruled out. 

The MF derived here differs from that obtained by \citet{dow08} who find power-law slopes of $\alpha\sim-1.8$ for $10^5 < M/M_\odot < 10^7$ for clusters in a sample of nearby spiral and irregular galaxies. The authors of this study point out that their cluster sample may be contaminated by blending and more diffuse associations because of the relatively poor resolution of their Sloan Digital Sky Survey data, which might lead to an overestimate of the number of very massive objects. They also note that completeness effects may have been underestimated at the low-mass end, which would lead to a too shallow overall slope. Interestingly, however, they find similar MFs in spirals and irregulars.

\section{Implications for the luminosity function}
\label{sec:lfs}

\begin{figure}
\centering
\includegraphics[width=85mm]{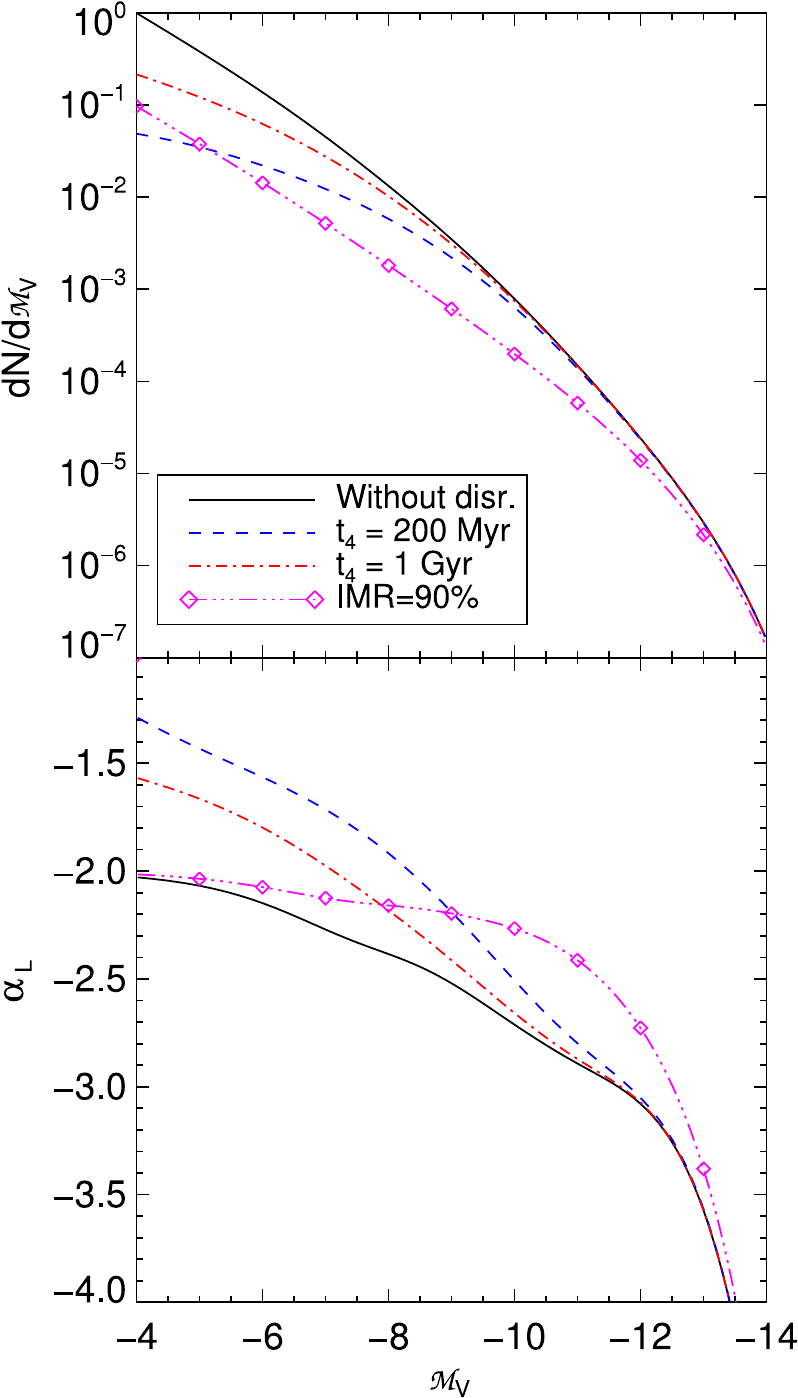}
\caption{\label{fig:lf}Top: luminosity functions for a cluster sample with a Schechter mass function with $\mbox{$M_c$}=2.1\times10^5\mbox{$M_{\odot}$}$. Bottom: local logarithmic slope. In all cases, a constant cluster formation history has been assumed and various disruption scenarios are applied, as indicated in the legend.
}
\end{figure}

Luminosity functions require no conversion from observed luminosities to masses via age-dependent mass-to-light ratios, so it is not surprising that more information is available about LFs than MFs. \citet{lar02} fitted power-laws $dN/dL \propto L^{\alpha_L}$ to the LFs of young  clusters in 6 spiral galaxies with HST/WFPC2 imaging, and found slopes in the range $-2 > \alpha_L > -2.4$ for typical $\mathcal{M}_V$ intervals $-6 > \mathcal{M}_V > -9$. A trend for the slopes to become steeper in brighter magnitude intervals was noted. Similar numbers were found for a sample of 5 spirals by \citet{mora08}, using HST/ACS data.  The most detailed analysis of the LF in a single galaxy is for NGC~5194, which has ACS imaging covering the entire disc. From these data, slopes between $\alpha_L=-2.5$ and $\alpha_L=-2.6$ have been obtained for clusters brighter than $\mathcal{M}_V=-8$ \citep{haas08,hl08}. This is similar to the LF slope derived by \citet{dk02} for clusters in NGC~3627: $\alpha_L = -2.53\pm0.15$ for $-8 > \mathcal{M}_V > -11$. \citet{gieles06b} have shown that the LF of clusters in NGC~5194 is consistent with clusters having been drawn from a Schechter mass function with $\mbox{$M_c$} \approx 10^5\mbox{$M_{\odot}$}$. 

The top panel in Fig.~\ref{fig:lf} shows the luminosity functions corresponding to various cluster disruption scenarios for a Schechter ICMF with $\mbox{$M_c$} = 2.1\times10^5\mbox{$M_{\odot}$}$ and $\alpha=-2$, plotted as number of clusters per magnitude bin. LFs are shown for no disruption, for secular dissolution with $t_4=200$ Myr  and $t_4=1$ Gyr, and for mass-independent disruption with IMR=90\% for $10^8$ years (i.e.\ the same as the Case C scenario in Sect.~\ref{sec:montecarlo}). These LFs were obtained by integrating  Eq.~(\ref{eq:dens}) over all $\tau$, assuming a constant cluster formation rate. It is evident that the LFs all steepen towards the bright end and hence cannot be well approximated by a single power-law. Local values of the logarithmic slope, i.e., the equivalent of the exponent of a local power-law fit $dN/dL \propto L^{\alpha_L}$, are plotted in the bottom panel of the figure. In the absence of secular, mass-dependent evolution the faint-end slope of the LF approaches that of the ICMF. At the bright end, the LF steepens, while at the faint end secular dissolution of low-mass clusters can lead to a shallower slope.

According to Fig.~\ref{fig:lf}, LF slopes steeper than $\alpha_L=-2$ are generally expected for the magnitude ranges typically covered in studies of extragalactic star cluster populations.  Comparing with the NGC~5194 observations specifically, the no-disruption model predicts a mean slope of $\langle \alpha_L \rangle = -2.53$ for $-8 > \mathcal{M}_V > -10$, in excellent agreement with the studies cited above. Inclusion of disruption effects leads to a somewhat shallower slope, e.g.\ $\langle \alpha_L \rangle = -2.20$ over the same $\mathcal{M}_V$ range for $t_4=2\times10^8$ years. However, age-dependent extinction can steepen the LF slope \citep{lar02} and there are indeed indications that younger clusters in M51 are more heavily extincted \citep{bast05}. Furthermore, the ACS mosaic for which the LF data were derived cover the whole disc where a longer disruption time scale than 200 Myr may be applicable, and the assumption of a constant cluster formation rate may well be invalid for M51. Lastly, the disruption time estimates for M51 were based on the assumption of an $\alpha=-2$ power-law ICMF. The smaller number of high-mass clusters predicted for an $\mbox{$M_c$}=2\times10^5\mbox{$M_{\odot}$}$ Schechter function would most strongly affect the detectable number of clusters at old ages, thus requiring less destruction and possibly lead to a longer estimate of the disruption time scale.
Interestingly, the faint end of the Milky Way open cluster LF ($-2 > \mathcal{M}_V > -9$) has a relatively shallow slope of $\alpha=-1.5$ \citep{vl84}. Due to secular dissolution, this is not necessarily inconsistent with an ICMF with low-mass slope $\alpha=-2$. In summary, a range of LF slopes is clearly expected when fitting cluster samples in different galaxies over different magnitude intervals. Generally, we conclude that an $\mbox{$M_c$}=2.1\times10^5\mbox{$M_{\odot}$}$ Schechter MF can lead to LFs that are consistent with observations of real cluster samples, but once again note that disruption and extinction effects make it difficult to draw strong conclusions about MFs based on observations of LFs alone.

\section{The cluster mass function in other environments - universal or not?}
\label{sec:elsewhere}

While the analysis in the previous sections indicates than an $\mbox{$M_c$}\approx2\times10^5\mbox{$M_{\odot}$}$ Schechter function fits the MFs of clusters in both cluster-poor and cluster-rich spirals (the latter also in different radial bins), the data are hardly sufficient to claim that this fit is universally valid. In the following, we examine to what extent this MF is consistent with observations of star clusters in other environments.

\subsection{Open clusters in the Milky Way}

When discussing the Milky Way open clusters, the poorly quantified completeness of current catalogues is a severe obstacle and the following comparison therefore remains necessarily sketchy. We first estimate, for a given cluster formation rate and MF, how many clusters should have formed in the Galactic disc in various mass intervals. We assume a star formation rate in bound clusters of $5.2\times10^{-10} \mbox{$M_{\odot}$}$ yr$^{-1}$ pc$^{-2}$ in the Solar neighbourhood \citepalias{lam05} and further approximate the Galactic disc as having a uniform cluster formation rate within a radius of $R=10$ kpc. The total cluster formation rate is then $\sim0.16$ \mbox{$M_{\odot}$} yr$^{-1}$, which we assume to have been constant in the past.  If this mass formed in clusters with masses sampled from a Schechter function with a lower mass limit of 100 \mbox{$M_{\odot}$} and $\mbox{$M_c$}=2\times10^5\mbox{$M_{\odot}$}$, the Milky Way has then formed about 19\,000 and 700 clusters more massive than $10^4\mbox{$M_{\odot}$}$ and $10^5\mbox{$M_{\odot}$}$ over a 10 Gyr period, respectively. This corresponds to one (bound) cluster more massive than $10^5 \mbox{$M_{\odot}$}$ forming about every 13 Myr somewhere in the disc, or one more massive than $10^4\mbox{$M_{\odot}$}$ about every $\sim500\,000$ years.  

About a dozen star clusters with ages $\la10$ Myr and masses in the range $10^4\mbox{$M_{\odot}$} - 10^5\mbox{$M_{\odot}$}$ have now been found within several kpc from the Sun \citep{figer08}.  Based on the above, only 5 clusters with $M>10^4\mbox{$M_{\odot}$}$ and younger than 10 Myr are expected within 5 kpc, but the calculation can hardly be considered more than an order-of-magnitude estimate and it is also possible that some of the young Milky Way clusters identified in current surveys will not remain bound. Hence, the observed numbers of young massive clusters does not seem to be in significant disagreement with what is expected based on the estimated formation rate of bound clusters and the assumption of an $\mbox{$M_c$}=2\times10^5\mbox{$M_{\odot}$}$ Schechter MF. 

Comparison with masses $10^5 - 10^6\mbox{$M_{\odot}$}$ is more difficult. No young disc clusters in this range have been identified, but only $\sim1$ is expected to have formed in the last $10^7$ years. About half of the total $\sim700$ clusters in this mass range would already have disrupted or have lost a large fraction of their initial mass, assuming that the \citetalias{lam05} disruption parameters were valid also in the past. If the remaining objects are evenly scattered across the Galactic disc, their average surface density is about 1 kpc$^{-2}$, so a few should be found within $\sim$1 kpc. This is in fair agreement with the estimate by \citetalias{lam05} \citep[based on the catalogue of][]{khar05} that the most massive cluster formed within 600 pc over the past few Gyr had an initial mass of $\sim10^5\mbox{$M_{\odot}$}$.

A 1-Gyr old cluster with an initial mass of $10^5 \mbox{$M_{\odot}$}$ should still have a present-day mass of $\sim$60\,000$\mbox{$M_{\odot}$}$. The surface density of clusters younger than 1 Gyr and initial masses above $10^5\mbox{$M_{\odot}$}$ should be about 0.25 kpc$^{-2}$ so a more complete survey of clusters in the Galactic disc out to several kpc might reveal a significant population of intermediate-age, massive clusters. It is of interest to note that the Milky Way does host a number of several Gyr-old open clusters which must initially have been quite massive \citep{jp94,friel95}. Identification of these objects is helped by the fact that many of them tend to be located quite far above the Galactic disc.

Extrapolating further to above $10^6 \mbox{$M_{\odot}$}$, the Milky Way disc would only form one such cluster about every 40 Gyr for an $\mbox{$M_c$}=2\times10^5\mbox{$M_{\odot}$}$ Schechter MF, hence it is not surprising that none has been found. This latter number is in stark contrast to the $\sim100$ clusters predicted by extrapolation of an $\alpha=-2$ power-law.

\subsection{The Large Magellanic Cloud}

\begin{figure}
\centering
\includegraphics[width=85mm]{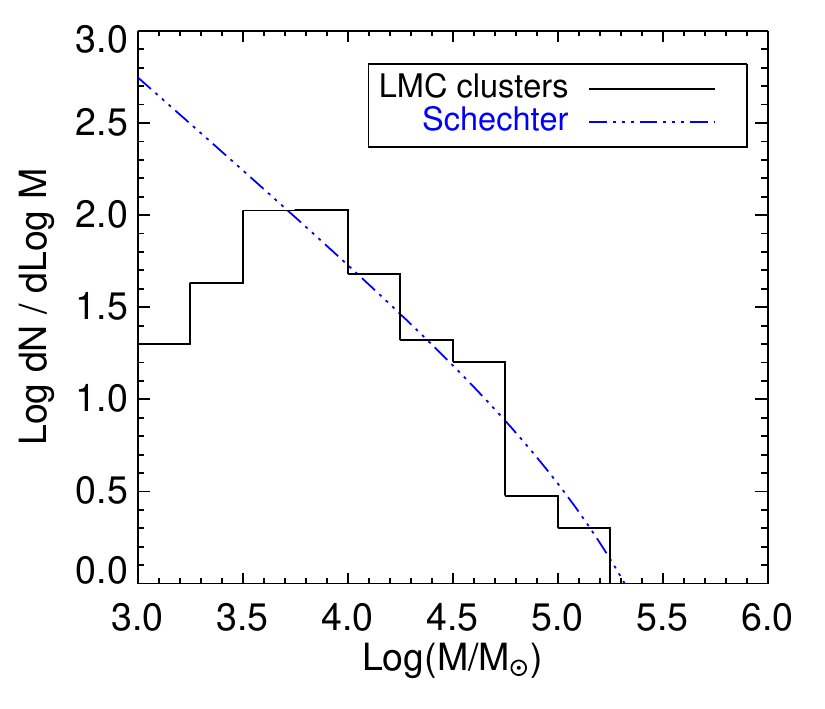}
\caption{\label{fig:mf_lmc}Mass function for star clusters in the LMC younger than 1 Gyr, based on photometry from \citet{bica96}. The dotted-dashed line is the same Schechter function as in Fig.~\ref{fig:mf_spir}, arbitrarily scaled.
}
\end{figure}

The LMC has long been known to host a rich population of young massive star clusters \citep[e.g.][]{hodge61}, and studies of the LMC clusters benefit from the advantage of having a clear view of the entire galaxy. For the brighter clusters, current databases are unlikely to suffer from significant incompleteness.  Excluding the old GCs, the most massive clusters in the LMC have masses of $\approx 2\times10^5\mbox{$M_{\odot}$}$ \citep{hunter03,and06}, and the MF can be well fitted by a power-law with slope $\alpha=-2$. Since the LMC cluster MF barely overlaps with the spiral data in Fig.~\ref{fig:mf_spir}, a direct comparison is not possible, but one can ask whether the LMC data are consistent with the Schechter MF derived for the spirals. To this end, the $UBV$ photometry tabulated by \citet{bica96} was used to derive ages and masses for 504 LMC clusters in the same way as was done for the spirals in Sect.~\ref{sec:mfs} (except that $Z=0.008$ SSP models were used, as appropriate for the LMC). A distance modulus of $m-\mathcal{M}=18.5$ and foreground extinction $A_V=0.25$ mag were assumed. 
The resulting MF for clusters younger than 1 Gyr is shown in Fig.~\ref{fig:mf_lmc} together with an $\mbox{$M_c$}=2.1\times10^5\mbox{$M_{\odot}$}$ Schechter function.  The \citet{bica96} data have an estimated completeness limit of $V\approx13$, which corresponds to a limiting mass of $\sim10^4\mbox{$M_{\odot}$}$ for an age of 1 Gyr. The more recent datasets go deeper, but for comparison with the spiral data the Bica et al.\ photometry suffices. This analysis yields 90 clusters more massive than $10^4\mbox{$M_{\odot}$}$, of which two exceed $10^5\mbox{$M_{\odot}$}$.  A K--S test comparing the LMC cluster MF for $M>10^4\mbox{$M_{\odot}$}$ with the Schechter function returns $P=0.94$, indicating excellent agreement (although the data are also consistent with an untruncated power-law; $P=0.56$). It can therefore be concluded that the LMC cluster MF can be fitted by the same MF derived for spirals, although there is little actual overlap in mass between the two samples. Thus, the impression that the LMC is particularly rich in ``massive'' clusters is most likely an observational selection effect. 

\subsection{Starbursts and mergers}

Turning to more extreme star forming environments, there are indications that the MF derived for spirals may not be adequate. We have already noted from the comparison of spiral and Antennae cluster MFs in Fig.~\ref{fig:mf_spir} that there are relatively more high-mass clusters in the Antennae sample. Indeed, a Schechter function fit to the Zhang \& Fall data yields $\log \mbox{$M_c$} = 5.9^{+0.45}_{-0.25}$ \citep{jordan07}, while an \mbox{$M_c$} that high is ruled out at high confidence level for the spirals. The conditions in the Antennae seem to be such that $\mbox{$M_c$}$ has shifted to a higher value, so that a relatively larger number of high-mass clusters form. The same appears to be the case in M82, although analysis of the cluster MF there is complicated by the large amount of extinction internally in the galaxy \citep{smith07}. From dynamical mass measurements for 15 clusters with masses between $2\times10^5\mbox{$M_{\odot}$}$ and $4\times10^6\mbox{$M_{\odot}$}$, \citet{mg07} fit a power-law with a slope of $\alpha=-1.91\pm0.06$. This is again inconsistent with an $\mbox{$M_c$}=2\times10^5\mbox{$M_{\odot}$}$ Schechter MF although selection effects remain difficult to quantify.

The mere presence of $>10^7\mbox{$M_{\odot}$}$ clusters in NGC~1316, NGC~7252 and Arp 220 \citep{bastian06,wilson06} is also inconsistent with a Schechter MF with an $\mbox{$M_c$}$ in the range $10^5\mbox{$M_{\odot}$} - 10^6\mbox{$M_{\odot}$}$. If a peak star formation rate of $\sim1000$ $\mbox{$M_{\odot}$}$ yr$^{-1}$ \citep{bastian08} were turned completely into clusters, the number of clusters with $M > 10^7\mbox{$M_{\odot}$}$ formed would be less than $5\times10^{-5}$ Myr$^{-1}$ for $\mbox{$M_c$} = 10^6 \mbox{$M_{\odot}$}$. Even if such a high SFR could be sustained for hundreds of Myrs, a negligible number of $10^7\mbox{$M_{\odot}$}$ clusters would form. Hence, it seems that the cluster MF in these more extreme environments must be different than that observed in spiral discs. 

\subsection{Old globular clusters}

Ultimately, it would be desirable to understand the properties of old GC populations in the context of their younger counterparts in present-day star forming galaxies. However, classical GCs have been subject to dynamical evolution over a Hubble time, which complicates the comparison. Few clusters with initial masses less than $\sim10^5\mbox{$M_{\odot}$}$ would have survived for a Hubble time \citep{vh97,bm03,fz01} and the shape of the low-mass end of the globular cluster MF is therefore strongly affected by dynamical evolution. At higher masses, the MF is more likely to resemble its original shape and there is substantial evidence of a steepening of the MF at the high-mass end in the rich GC systems of giant elliptical galaxies \citep{mp96,bs00,jordan07}. The ``evolved Schechter function'' fits to mass functions for GC systems in Virgo cluster galaxies by  \citet{jordan07} indicate an $\mbox{$M_c$}$ increasing steadily from $(3-4)\times10^5\mbox{$M_{\odot}$}$ in the faintest galaxies ($\mathcal{M}_V \sim -16$) to $(2-3)\times10^6\mbox{$M_{\odot}$}$ in the giant ellipticals. Given such evidence of variations in \mbox{$M_c$} even among old GC systems, it is not surprising that the present-day ICMF may also be environmentally dependent. Ultimately, understanding these variations may be a powerful tool for constraining the formation conditions for GC systems at cosmological epochs.

\section{Discussion}
\label{sec:discussion}

The suggestion that the ICMF has a characteristic upper mass scale naturally raises the question as to what physical mechanism is responsible. It is tempting to assume that the ICMF is related to the mass function of the giant molecular clouds (GMCs) in which cluster formation takes place. The MF of GMCs in the Milky Way and other Local Group galaxies can be fitted by a power-law with slope $\approx-1.7$, slightly shallower than for the cluster MF, below a fairly well defined upper limit of $M \approx 6\times10^6\mbox{$M_{\odot}$}$ \citep{wmk97,blitz07,roso07}. Much of the mass in a (Galactic) GMC is at low density and global star formation efficiencies (SFEs) are typically only a few percent \citep{myers86,ll03}. Although it is not necessarily true that all star formation in a GMC results in a single cluster, the upper GMC mass limit and low SFEs seem roughly compatible with the $\mbox{$M_c$}$ derived in this paper for the ICMF in spiral galaxies. However, the \emph{local} SFE within cluster-forming clumps in GMCs has to be high to produce a bound star cluster \citep[30\%--50\%;][]{hills80,elm83,bk07} and ultimately the mass spectrum of these clumps may be more relevant.

The formation of molecular clouds and the origin of their mass function remain poorly understood. The low-mass slope may follow from fractal structure in a turbulence-dominated interstellar medium (ISM) \citep{ef96}. It has often been suggested that the upper mass limit is related to the maximum unstable length scale in a rotating disc, due to shear and/or the coriolis force \citep{toomre64,heb98,dow08}. Analytic arguments and detailed numerical simulations lead to maximum cloud masses of $\sim10^7\mbox{$M_{\odot}$}$, in rough agreement with observations \citep{mo07}. The maximum cloud mass may be expected to scale with gas/total mass fraction and total gas mass in discs, so that more massive clouds should be able to form e.g.\ in gas-rich proto-galactic discs \citep{el08}. However, the picture is complicated by the highly dynamic character of the interstellar medium. GMCs are probably rather transient structures \citep{mk04,elme07} so it is unclear to what extent an overall stability criterion is applicable in general.

What might be responsible for shifting the maximum cluster mass upwards in starburst environments? If the most massive clusters form at similarly low global efficiencies, massive ``Super-GMCs'' (SGMCs) would be required \citep{hp94,mp96}. In the context of old GC systems, these SGMCs may be identified with \citet{sz78}-style fragments and cosmological simulations suggest that the first GCs may indeed have formed in such clouds, embedded within dark matter halos \citep{kg05}. Some evidence that elements of this scenario may be relevant in present-day star forming environments is provided by observations of cloud complexes with masses up to nearly $10^9\mbox{$M_{\odot}$}$ in the Antennae \citep{wilson03}. However, the Antennae is too far away to study these structures in much detail with current radio interferometers. Giant molecular associations with masses of $10^7 - 10^8$\mbox{$M_{\odot}$}, potentially suitable for forming $10^5-10^6\mbox{$M_{\odot}$}$ clusters at low overall efficiency, have been found in nearby cluster-rich spirals like NGC~5194 and NGC~5236 but tend to break up into smaller GMC-like structures when observed at high resolution \citep{vogel88,kuno95,rand99,lund04}. In general, the ISM in galaxies is hierarchically structured also on scales above the scale of individual GMCs \citep{ef96,elme07} and SGMC-like structures may appear conspicuous in the Antennae simply because a larger fraction of the gas is in molecular form there \citep{gao01}. In \object{M82}, the nearest starburst, observations at high spatial resolution have failed to reveal distinct SGMCs \citep{keto05}.

Alternatively, a shift towards higher maximum cluster mass could result if the global SFE within more modest-sized GMCs is increased. In mergers, GMCs may be compressed to high densities by an increased ambient ISM pressure as the two galaxies collide, leading to a strong increase in the overall star formation rate and formation of massive, bound clusters \citep{js92,az01}. Indeed, observations show that a large fraction of the interstellar medium in starburst galaxies (many of which are also mergers) is in a dense phase \citep{gs04}. There is some evidence of shock-triggered cluster formation in the Antennae, but large-scale shocks are probably not directly responsible for most cluster formation there \citep{zfw01}. GMCs in M82 also seem to be shock-compressed, in this case not due to a major merger but due to the pressure from surrounding ionized gas \citep{keto05}, a mechanism which may also be at work in the Antennae.

The distinction between formation of massive clusters in dense clumps within SGMCs and in normal-sized but compressed GMCs may be somewhat artificial in most present-day environments, since any dense concentration of molecular gas will always be surrounded by lower-density material. A more relevant issue may be what delivers the pressure necessary to confine a large amount of gas in dense clumps while star formation is taking place \citep{ee97}. Here, differences in local conditions of the interstellar medium, such as gas kinematics, may be of relevance. The most actively star-forming spiral galaxies in our sample, like NGC~6946, have overall star formation rates (SFRs) which are no more than a factor of a few smaller than in the Antennae \citep{lar02,sauty98,stan90,zfw01}. In spite of this high SFR, the kinematic situation in NGC~6946 (as in most spirals) is relatively quiescent. The gas resides in a large, orderly rotating disc \citep{boom08} with no major distortions apart from a number of (probably extra-planar) high velocity clouds. The galaxy is among the most isolated in the nearby Universe \citep{pw00}, so its high SFR is unlikely to be interaction-induced. The gas kinematics in the Antennae are instead complex with a velocity spread of several hundred km s$^{-1}$, notably in the overlap region of the two galaxies where star formation is currently most active \citep{gao01,gordon01,gg07}.  The starburst in M82 is driving a superwind with outflow velocities in excess of 100 km s$^{-1}$, carrying a significant fraction of the molecular gas in the system \citep{walter02}, again showing that a large amount of energy is being injected into the interstellar gas.  Similar complex gas kinematics are observed in other starbursts \citep{arri01,martin06}. In these environments, formation of massive clusters may be favoured in regions of high gas pressure \citep{ee97,keto05}. Theoretically, this is supported by simulations of major mergers \citep{mh96,li04,bour08} which show that a larger fraction of the gas in the mergers is found in regions of high pressure and high velocity dispersion ($>30$ km s$^{-1}$) compared to a spiral disc. 

It is difficult to tell whether star formation in spirals and mergers/starbursts represent truly different modes or are part of a continuum. While a relatively larger fraction of the gas in violent starbursts is in high-pressure regions, such conditions may also be encountered more locally in spiral discs \citep{eel00,efre07}. Indeed, the comparison in Fig.~\ref{fig:mf_spir} shows that the maximum cluster mass encountered in spirals is not very different from that in the Antennae, although the high-mass slope of the ICMF appears steeper in the spirals.  The Antennae is still in an early stage of the merging process and the conditions are likely to get more extreme over the next several $10^8$ years \citep{mbr93} with a possible shift of \mbox{$M_c$} to even higher values.  There may well be a range of possible \mbox{$M_c$} values in present-day star forming environments and not just two distinct (``quiescent'' and ``starburst'') modes.

Some of the most massive clusters might form by a different mechanism altogether, such as merging of less massive clusters \citep{eel00,fk05}. This would bypass the need for a mechanism to collect $>10^7\mbox{$M_{\odot}$}$ of gas in a single dense clump, but would still require many lower-mass clusters to form sufficiently near each other to merge. The merger products are predicted to be relatively diffuse so this mechanism is unlikely to account for more compact clusters such as those found in \object{NGC~1316} and \object{NGC~7252} \citep{bastian06}. However, extremely massive objects, like the $8\times10^7\mbox{$M_{\odot}$}$ object W3 in NGC~7252 \citep{mar04}, could have formed this way.

\section{Summary and conclusions}

This paper provides three main lines of evidence suggesting that the mass function of young star clusters in spiral galaxies can be approximated by a Schechter function with $\mbox{$M_c$}=2.1\times10^5 \mbox{$M_{\odot}$}$ and low-mass slope $\alpha=-2$:
\begin{itemize}
  \item The observed age-luminosity relations for the brightest and \mbox{$5^{\rm th}$} brightest cluster in a sample of mostly late-type spirals are consistent with those expected for cluster populations drawn from a Schechter MF with an \mbox{$M_c$} of a few times $10^5\mbox{$M_{\odot}$}$. This result is not significantly affected by secular cluster disruption.
  \item A maximum-likelihood fit of a Schechter function to the combined MF for clusters more massive than $10^5\mbox{$M_{\odot}$}$ in these spirals yields $\mbox{$M_c$}=(2.1\pm0.4)\times10^5\mbox{$M_{\odot}$}$ for fixed low-mass slope $\alpha=-2$. The cluster MFs for cluster-poor and cluster-rich subsamples of the galaxies are statistically indistinguishable.
  \item The luminosity function for a model cluster sample with masses drawn from an $\mbox{$M_c$}=2.1\times10^5\mbox{$M_{\odot}$}$ Schechter function is consistent with existing data for the LFs of clusters in spiral galaxies, including the well studied M51 cluster system.
\end{itemize}
An $\mbox{$M_c$}=2.1\times10^5\mbox{$M_{\odot}$}$ (or $\mbox{$M_c$}\approx1.4\times10^5\mbox{$M_{\odot}$}$ for a Kroupa- or Chabrier IMF) Schechter function also fits the MF of clusters in the Large Magellanic Cloud, and it is argued that it is probably consistent with current surveys of young clusters in the Milky Way. It is, however, unlikely to be universal. In merger galaxies and other starbursts, a Schechter function fit requires a significantly higher \mbox{$M_c$} than derived for the spirals. Since some spirals are not very different from mergers like the Antennae in terms of overall star formation rate or gas content, differences in the upper end of the cluster MFs are more likely related to the different gas kinematics in these systems. The fact that \mbox{$M_c$} values are found to be greater for old GC systems than for young cluster populations in discs
suggests that star (cluster) formation in present-day spiral discs generally proceeds in a relatively more orderly way than when most old GCs formed. In the future, observations with ALMA should provide crucial insight into the properties of molecular gas on scales that are directly relevant to formation of individual clusters in galaxies well beyond the Local Group.

\begin{acknowledgements}
I thank Mark Gieles for encouragement to get this work published and Brad Whitmore for an interesting discussion during the \emph{Science with the new Hubble Space Telescope} meeting in Bologna, January 2008.  The manuscript has benefited from helpful comments from Mark Gieles, Henny Lamers and the referee, Nate Bastian.
\end{acknowledgements}

\bibliographystyle{aa}
\bibliography{refs.bib}

\begin{thebibliography}{129}
\expandafter\ifx\csname natexlab\endcsname\relax\def\natexlab#1{#1}\fi

\bibitem[{Anders \& de~Grijs(2006)}]{and06}
Anders, P. \& de~Grijs, R. 2006, \mnras, 366, 295

\bibitem[{Anders {et~al.}(2004)Anders, de~Grijs, {Fritze-v.~Alvensleben}, \&
  Bissantz}]{and04}
Anders, P., de~Grijs, R., {Fritze-v.~Alvensleben}, U., \& Bissantz, N. 2004,
  \mnras, 347, 17

\bibitem[{{Angeretti} {et~al.}(2005){Angeretti}, {Tosi}, {Greggio}, {Sabbi},
  {Aloisi}, \& {Leitherer}}]{ange05}
{Angeretti}, L., {Tosi}, M., {Greggio}, L., {et~al.} 2005, \aj, 129, 2203

\bibitem[{{Annibali} {et~al.}(2003){Annibali}, {Greggio}, {Tosi}, {Aloisi}, \&
  {Leitherer}}]{anni03}
{Annibali}, F., {Greggio}, L., {Tosi}, M., {Aloisi}, A., \& {Leitherer}, C.
  2003, \aj, 126, 2752

\bibitem[{{Arribas} {et~al.}(2001){Arribas}, {Colina}, \& {Clements}}]{arri01}
{Arribas}, S., {Colina}, L., \& {Clements}, D. 2001, \apj, 560, 160

\bibitem[{{Ashman} \& {Zepf}(2001)}]{az01}
{Ashman}, K.~M. \& {Zepf}, S.~E. 2001, \aj, 122, 1888

\bibitem[{{Bastian}(2008)}]{bastian08}
{Bastian}, N. 2008, \mnras, 390, 759

\bibitem[{Bastian {et~al.}(2005)Bastian, Gieles, Lamers, Scheepmaker, \&
  de~Grijs}]{bast05}
Bastian, N., Gieles, M., Lamers, H. J. G. L.~M., Scheepmaker, R.~A., \&
  de~Grijs, R. 2005, \aap, 431, 905

\bibitem[{{Bastian} {et~al.}(2006){Bastian}, {Saglia}, {Goudfrooij},
  {Kissler-Patig}, {Maraston}, {Schweizer}, \& {Zoccali}}]{bastian06}
{Bastian}, N., {Saglia}, R.~P., {Goudfrooij}, P., {et~al.} 2006, \aap, 448, 881

\bibitem[{Baumgardt \& Kroupa(2007)}]{bk07}
Baumgardt, H. \& Kroupa, P. 2007, \mnras, 380, 1589

\bibitem[{Baumgardt \& Makino(2003)}]{bm03}
Baumgardt, H. \& Makino, J. 2003, \mnras, 340, 227

\bibitem[{{Bica} {et~al.}(1996){Bica}, {Claria}, {Dottori}, {Santos}, \&
  {Piatti}}]{bica96}
{Bica}, E., {Claria}, J.~J., {Dottori}, H., {Santos}, Jr., J.~F.~C., \&
  {Piatti}, A.~E. 1996, \apjs, 102, 57

\bibitem[{{Bik} {et~al.}(2003){Bik}, {Lamers}, {Bastian}, {Panagia}, \&
  {Romaniello}}]{bik03}
{Bik}, A., {Lamers}, H.~J.~G.~L.~M., {Bastian}, N., {Panagia}, N., \&
  {Romaniello}, M. 2003, \aap, 397, 473

\bibitem[{{Billett} {et~al.}(2002){Billett}, {Hunter}, \& {Elmegreen}}]{bhe02}
{Billett}, O.~H., {Hunter}, D.~A., \& {Elmegreen}, B.~G. 2002, \aj, 123, 1454

\bibitem[{{Blitz} {et~al.}(2007){Blitz}, {Fukui}, {Kawamura}, {Leroy},
  {Mizuno}, \& {Rosolowsky}}]{blitz07}
{Blitz}, L., {Fukui}, Y., {Kawamura}, A., {et~al.} 2007, in Protostars and
  Planets V, ed. B.~{Reipurth}, D.~{Jewitt}, \& K.~{Keil}, 81--96

\bibitem[{{Boomsma} {et~al.}(2008){Boomsma}, {Oosterloo}, {Fraternali}, {van
  der Hulst}, \& {Sancisi}}]{boom08}
{Boomsma}, R., {Oosterloo}, T.~A., {Fraternali}, F., {van der Hulst}, J.~M., \&
  {Sancisi}, R. 2008, \aap, 490, 555

\bibitem[{{Bournaud} {et~al.}(2008){Bournaud}, {Duc}, \& {Emsellem}}]{bour08}
{Bournaud}, F., {Duc}, P.-A., \& {Emsellem}, E. 2008, \mnras, 389, L8

\bibitem[{Brodie \& Strader(2006)}]{bs06}
Brodie, J.~P. \& Strader, J. 2006, \araa, 44, 193

\bibitem[{{Bruzual} \& {Charlot}(2003)}]{bc03}
{Bruzual}, G. \& {Charlot}, S. 2003, \mnras, 344, 1000

\bibitem[{{Burkert} \& {Smith}(2000)}]{bs00}
{Burkert}, A. \& {Smith}, G.~H. 2000, \apjl, 542, L95

\bibitem[{{Chabrier}(2003)}]{chab03}
{Chabrier}, G. 2003, \pasp, 115, 763

\bibitem[{{Chandar} {et~al.}(2006){Chandar}, {Fall}, \& {Whitmore}}]{cfw06}
{Chandar}, R., {Fall}, S.~M., \& {Whitmore}, B.~C. 2006, \apjl, 650, L111

\bibitem[{{de Grijs} {et~al.}(2003){de Grijs}, {Anders}, {Bastian}, {Lynds},
  {Lamers}, \& {O'Neil}}]{deg03}
{de Grijs}, R., {Anders}, P., {Bastian}, N., {et~al.} 2003, \mnras, 343, 1285

\bibitem[{{de Grijs} \& {Goodwin}(2008)}]{deg08}
{de Grijs}, R. \& {Goodwin}, S.~P. 2008, \mnras, 383, 1000

\bibitem[{{de Grijs} \& {Parmentier}(2007)}]{dp07}
{de Grijs}, R. \& {Parmentier}, G. 2007, Chinese Journal of Astronomy and
  Astrophysics, 7, 155

\bibitem[{{Dolphin} \& {Kennicutt}(2002)}]{dk02}
{Dolphin}, A.~E. \& {Kennicutt}, Jr., R.~C. 2002, \aj, 123, 207

\bibitem[{{Dowell} {et~al.}(2008){Dowell}, {Buckalew}, \& {Tan}}]{dow08}
{Dowell}, J.~D., {Buckalew}, B.~A., \& {Tan}, J.~C. 2008, \aj, 135, 823

\bibitem[{{Efremov} {et~al.}(2007){Efremov}, {Afanasiev}, {Alfaro}, {Boomsma},
  {Bastian}, {Larsen}, {S{\'a}nchez-Gil}, {Silchenko}, {Garc{\'{\i}}a-Lorenzo},
  {Mu{\~n}oz-Tu{\~n}on}, \& {Hodge}}]{efre07}
{Efremov}, Y.~N., {Afanasiev}, V.~L., {Alfaro}, E.~J., {et~al.} 2007, \mnras,
  382, 481

\bibitem[{{Elmegreen}(1983)}]{elm83}
{Elmegreen}, B.~G. 1983, \mnras, 203, 1011

\bibitem[{{Elmegreen}(2007)}]{elme07}
{Elmegreen}, B.~G. 2007, \apj, 668, 1064

\bibitem[{{Elmegreen} \& {Efremov}(1997)}]{ee97}
{Elmegreen}, B.~G. \& {Efremov}, Y.~N. 1997, \apj, 480, 235

\bibitem[{{Elmegreen} {et~al.}(2000){Elmegreen}, {Efremov}, \&
  {Larsen}}]{eel00}
{Elmegreen}, B.~G., {Efremov}, Y.~N., \& {Larsen}, S. 2000, \apj, 535, 748

\bibitem[{{Elmegreen} \& {Falgarone}(1996)}]{ef96}
{Elmegreen}, B.~G. \& {Falgarone}, E. 1996, \apj, 471, 816

\bibitem[{{Escala} \& {Larson}(2008)}]{el08}
{Escala}, A. \& {Larson}, R.~B. 2008, \apjl, 685, L31

\bibitem[{{Fall}(2006)}]{fall06}
{Fall}, S.~M. 2006, \apj, 652, 1129

\bibitem[{{Fall} {et~al.}(2005){Fall}, {Chandar}, \& {Whitmore}}]{fcw05}
{Fall}, S.~M., {Chandar}, R., \& {Whitmore}, B.~C. 2005, \apjl, 631, L133

\bibitem[{{Fall} \& {Zhang}(2001)}]{fz01}
{Fall}, S.~M. \& {Zhang}, Q. 2001, \apj, 561, 751

\bibitem[{{Fellhauer} \& {Kroupa}(2005)}]{fk05}
{Fellhauer}, M. \& {Kroupa}, P. 2005, \mnras, 359, 223

\bibitem[{{Figer}(2008)}]{figer08}
{Figer}, D.~F. 2008, in IAU Symposium, Vol. 250, IAU Symposium, 247--256

\bibitem[{{Friel}(1995)}]{friel95}
{Friel}, E.~D. 1995, \araa, 33, 381

\bibitem[{{Fuchs} {et~al.}(2008){Fuchs}, {Jahrei\ss}, \& {Flynn}}]{fuchs08}
{Fuchs}, B., {Jahrei\ss}, H., \& {Flynn}, C. 2008, \aj, in press (preprint
  astro-ph/0810.1656)

\bibitem[{{Gao} {et~al.}(2001){Gao}, {Lo}, {Lee}, \& {Lee}}]{gao01}
{Gao}, Y., {Lo}, K.~Y., {Lee}, S.-W., \& {Lee}, T.-H. 2001, \apj, 548, 172

\bibitem[{{Gao} \& {Solomon}(2004)}]{gs04}
{Gao}, Y. \& {Solomon}, P.~M. 2004, \apj, 606, 271

\bibitem[{{Gelatt} {et~al.}(2001){Gelatt}, {Hunter}, \& {Gallagher}}]{gel01}
{Gelatt}, A.~E., {Hunter}, D.~A., \& {Gallagher}, J.~S. 2001, \pasp, 113, 142

\bibitem[{{Gieles} \& {Bastian}(2008)}]{gibast08}
{Gieles}, M. \& {Bastian}, N. 2008, \aap, 482, 165

\bibitem[{{Gieles} {et~al.}(2005){Gieles}, {Bastian}, {Lamers}, \&
  {Mout}}]{gieles05}
{Gieles}, M., {Bastian}, N., {Lamers}, H.~J.~G.~L.~M., \& {Mout}, J.~N. 2005,
  \aap, 441, 949

\bibitem[{{Gieles} \& {Baumgardt}(2008)}]{gb08}
{Gieles}, M. \& {Baumgardt}, H. 2008, \mnras, 389, L28

\bibitem[{{Gieles} {et~al.}(2007){Gieles}, {Lamers}, \& {Portegies
  Zwart}}]{gieles07b}
{Gieles}, M., {Lamers}, H.~J.~G.~L.~M., \& {Portegies Zwart}, S.~F. 2007, \apj,
  668, 268

\bibitem[{{Gieles} {et~al.}(2006{\natexlab{a}}){Gieles}, {Larsen}, {Bastian},
  \& {Stein}}]{gieles06}
{Gieles}, M., {Larsen}, S.~S., {Bastian}, N., \& {Stein}, I.~T.
  2006{\natexlab{a}}, \aap, 450, 129

\bibitem[{{Gieles} {et~al.}(2006{\natexlab{b}}){Gieles}, {Larsen},
  {Scheepmaker}, {Bastian}, {Haas}, \& {Lamers}}]{gieles06b}
{Gieles}, M., {Larsen}, S.~S., {Scheepmaker}, R.~A., {et~al.}
  2006{\natexlab{b}}, \aap, 446, L9

\bibitem[{{Gilbert} \& {Graham}(2007)}]{gg07}
{Gilbert}, A.~M. \& {Graham}, J.~R. 2007, \apj, 668, 168

\bibitem[{{Girardi} {et~al.}(1995){Girardi}, {Chiosi}, {Bertelli}, \&
  {Bressan}}]{gir95}
{Girardi}, L., {Chiosi}, C., {Bertelli}, G., \& {Bressan}, A. 1995, \aap, 298,
  87

\bibitem[{{Goodwin} \& {Bastian}(2006)}]{gb06}
{Goodwin}, S.~P. \& {Bastian}, N. 2006, \mnras, 373, 752

\bibitem[{{Gordon} {et~al.}(2001){Gordon}, {Koribalski}, \& {Jones}}]{gordon01}
{Gordon}, S., {Koribalski}, B., \& {Jones}, K. 2001, \mnras, 326, 578

\bibitem[{{Haas} {et~al.}(2008){Haas}, {Gieles}, {Scheepmaker}, {Larsen}, \&
  {Lamers}}]{haas08}
{Haas}, M.~R., {Gieles}, M., {Scheepmaker}, R.~A., {Larsen}, S.~S., \&
  {Lamers}, H.~J.~G.~L.~M. 2008, \aap, 487, 937

\bibitem[{{Harris} \& {Pudritz}(1994)}]{hp94}
{Harris}, W.~E. \& {Pudritz}, R.~E. 1994, \apj, 429, 177

\bibitem[{{Hernandez} {et~al.}(2000){Hernandez}, {Valls-Gabaud}, \&
  {Gilmore}}]{her00}
{Hernandez}, X., {Valls-Gabaud}, D., \& {Gilmore}, G. 2000, \mnras, 316, 605

\bibitem[{{Hills}(1980)}]{hills80}
{Hills}, J.~G. 1980, \apj, 235, 986

\bibitem[{{Ho} \& {Filippenko}(1996)}]{hf96}
{Ho}, L.~C. \& {Filippenko}, A.~V. 1996, \apjl, 466, L83

\bibitem[{{Hodge}(1961)}]{hodge61}
{Hodge}, P.~W. 1961, \apj, 133, 413

\bibitem[{{Hunter} {et~al.}(1998){Hunter}, {Elmegreen}, \& {Baker}}]{heb98}
{Hunter}, D.~A., {Elmegreen}, B.~G., \& {Baker}, A.~L. 1998, \apj, 493, 595

\bibitem[{{Hunter} {et~al.}(2003){Hunter}, {Elmegreen}, {Dupuy}, \&
  {Mortonson}}]{hunter03}
{Hunter}, D.~A., {Elmegreen}, B.~G., {Dupuy}, T.~J., \& {Mortonson}, M. 2003,
  \aj, 126, 1836

\bibitem[{{Hwang} \& {Lee}(2008)}]{hl08}
{Hwang}, N. \& {Lee}, M.~G. 2008, \aj, 135, 1567

\bibitem[{{Janes} \& {Phelps}(1994)}]{jp94}
{Janes}, K.~A. \& {Phelps}, R.~L. 1994, \aj, 108, 1773

\bibitem[{{Jog} \& {Solomon}(1992)}]{js92}
{Jog}, C.~J. \& {Solomon}, P.~M. 1992, \apj, 387, 152

\bibitem[{{Jord{\'a}n} {et~al.}(2007){Jord{\'a}n}, {McLaughlin},
  {C{\^o}t{\'e}}, {Ferrarese}, {Peng}, {Mei}, {Villegas}, {Merritt}, {Tonry},
  \& {West}}]{jordan07}
{Jord{\'a}n}, A., {McLaughlin}, D.~E., {C{\^o}t{\'e}}, P., {et~al.} 2007,
  \apjs, 171, 101

\bibitem[{{Keto} {et~al.}(2005){Keto}, {Ho}, \& {Lo}}]{keto05}
{Keto}, E., {Ho}, L.~C., \& {Lo}, K.-Y. 2005, \apj, 635, 1062

\bibitem[{{Kharchenko} {et~al.}(2005){Kharchenko}, {Piskunov}, {R{\"o}ser},
  {Schilbach}, \& {Scholz}}]{khar05}
{Kharchenko}, N.~V., {Piskunov}, A.~E., {R{\"o}ser}, S., {Schilbach}, E., \&
  {Scholz}, R.-D. 2005, \aap, 438, 1163

\bibitem[{{Kravtsov} \& {Gnedin}(2005)}]{kg05}
{Kravtsov}, A.~V. \& {Gnedin}, O.~Y. 2005, \apj, 623, 650

\bibitem[{{Kroupa}(2002)}]{kroupa02}
{Kroupa}, P. 2002, Science, 295, 82

\bibitem[{{Kroupa} \& {Boily}(2002)}]{kb02}
{Kroupa}, P. \& {Boily}, C.~M. 2002, \mnras, 336, 1188

\bibitem[{{Kruijssen}(2008)}]{kruij08}
{Kruijssen}, J.~M.~D. 2008, \aap, 486, L21

\bibitem[{{Kuno} {et~al.}(1995){Kuno}, {Nakai}, {Handa}, \& {Sofue}}]{kuno95}
{Kuno}, N., {Nakai}, N., {Handa}, T., \& {Sofue}, Y. 1995, \pasj, 47, 745

\bibitem[{Lada \& Lada(2003)}]{ll03}
Lada, C.~J. \& Lada, E.~A. 2003, \araa, 41, 57

\bibitem[{{Lamers} \& {Gieles}(2008)}]{lg08}
{Lamers}, H.~J.~G.~L.~M. \& {Gieles}, M. 2008, in Astronomical Society of the
  Pacific Conference Series, Vol. 388, Mass Loss from Stars and the Evolution
  of Stellar Clusters, ed. A.~{de Koter}, L.~J. {Smith}, \& L.~B.~F.~M.
  {Waters}, 367

\bibitem[{{Lamers} {et~al.}(2005){Lamers}, {Gieles}, {Bastian}, {Baumgardt},
  {Kharchenko}, \& {Portegies Zwart}}]{lam05}
{Lamers}, H.~J.~G.~L.~M., {Gieles}, M., {Bastian}, N., {et~al.} 2005, \aap,
  441, 117

\bibitem[{{Larsen}(1999)}]{lar99}
{Larsen}, S.~S. 1999, \aaps, 139, 393

\bibitem[{{Larsen}(2002)}]{lar02}
{Larsen}, S.~S. 2002, \aj, 124, 1393

\bibitem[{{Larsen}(2006)}]{lar04}
{Larsen}, S.~S. 2006, in Planets to Cosmology: Essential Science in the Final
  Years of the Hubble Space Telescope, ed. M.~{Livio} \& S.~{Casertano}, 35

\bibitem[{{Larsen} {et~al.}(2001){Larsen}, {Brodie}, {Elmegreen}, {Efremov},
  {Hodge}, \& {Richtler}}]{lar01}
{Larsen}, S.~S., {Brodie}, J.~P., {Elmegreen}, B.~G., {et~al.} 2001, \apj, 556,
  801

\bibitem[{{Larsen} {et~al.}(2004){Larsen}, {Brodie}, \& {Hunter}}]{lbh04}
{Larsen}, S.~S., {Brodie}, J.~P., \& {Hunter}, D.~A. 2004, \aj, 128, 2295

\bibitem[{{Larsen} \& {Richtler}(2000)}]{lr00}
{Larsen}, S.~S. \& {Richtler}, T. 2000, \aap, 354, 836

\bibitem[{{Larsen} \& {Richtler}(2004)}]{lr04}
{Larsen}, S.~S. \& {Richtler}, T. 2004, \aap, 427, 495

\bibitem[{{Li} {et~al.}(2004){Li}, {Mac Low}, \& {Klessen}}]{li04}
{Li}, Y., {Mac Low}, M.-M., \& {Klessen}, R.~S. 2004, \apjl, 614, L29

\bibitem[{{Lundgren} {et~al.}(2004){Lundgren}, {Wiklind}, {Olofsson}, \&
  {Rydbeck}}]{lund04}
{Lundgren}, A.~A., {Wiklind}, T., {Olofsson}, H., \& {Rydbeck}, G. 2004, \aap,
  413, 505

\bibitem[{{Mac Low} \& {Klessen}(2004)}]{mk04}
{Mac Low}, M.-M. \& {Klessen}, R.~S. 2004, Reviews of Modern Physics, 76, 125

\bibitem[{Maeder \& Mermilliod(1981)}]{mm81}
Maeder, A. \& Mermilliod, J.-C. 1981, \aap, 93, 136

\bibitem[{{Maraston} {et~al.}(2004){Maraston}, {Bastian}, {Saglia},
  {Kissler-Patig}, {Schweizer}, \& {Goudfrooij}}]{mar04}
{Maraston}, C., {Bastian}, N., {Saglia}, R.~P., {et~al.} 2004, \aap, 416, 467

\bibitem[{{Martin}(2006)}]{martin06}
{Martin}, C.~L. 2006, \apj, 647, 222

\bibitem[{{McCrady} \& {Graham}(2007)}]{mg07}
{McCrady}, N. \& {Graham}, J.~R. 2007, \apj, 663, 844

\bibitem[{{McKee} \& {Ostriker}(2007)}]{mo07}
{McKee}, C.~F. \& {Ostriker}, E.~C. 2007, \araa, 45, 565

\bibitem[{{McLaughlin} \& {Fall}(2008)}]{mf08}
{McLaughlin}, D.~E. \& {Fall}, S.~M. 2008, \apj, 679, 1272

\bibitem[{{McLaughlin} \& {Pudritz}(1996)}]{mp96}
{McLaughlin}, D.~E. \& {Pudritz}, R.~E. 1996, \apj, 457, 578

\bibitem[{{Mihos} {et~al.}(1993){Mihos}, {Bothun}, \& {Richstone}}]{mbr93}
{Mihos}, J.~C., {Bothun}, G.~D., \& {Richstone}, D.~O. 1993, \apj, 418, 82

\bibitem[{{Mihos} \& {Hernquist}(1996)}]{mh96}
{Mihos}, J.~C. \& {Hernquist}, L. 1996, \apj, 464, 641

\bibitem[{{Mora} {et~al.}(2008){Mora}, {Larsen}, {Kissler-Patig}, {Brodie}, \&
  {Richtler}}]{mora08}
{Mora}, M.~D., {Larsen}, S.~S., {Kissler-Patig}, M., {Brodie}, J.~P., \&
  {Richtler}, T. 2008, \aap, submitted

\bibitem[{{Myers} {et~al.}(1986){Myers}, {Dame}, {Thaddeus}, {Cohen},
  {Silverberg}, {Dwek}, \& {Hauser}}]{myers86}
{Myers}, P.~C., {Dame}, T.~M., {Thaddeus}, P., {et~al.} 1986, \apj, 301, 398

\bibitem[{{Parmentier} \& {Fritze}(2008)}]{pf08}
{Parmentier}, G. \& {Fritze}, U. 2008, \apj, in press (preprint
  astro-ph/0809.2416)

\bibitem[{{Pisano} \& {Wilcots}(2000)}]{pw00}
{Pisano}, D.~J. \& {Wilcots}, E.~M. 2000, \mnras, 319, 821

\bibitem[{{Rand} {et~al.}(1999){Rand}, {Lord}, \& {Higdon}}]{rand99}
{Rand}, R.~J., {Lord}, S.~D., \& {Higdon}, J.~L. 1999, \apj, 513, 720

\bibitem[{{Rocha-Pinto} {et~al.}(2000){Rocha-Pinto}, {Scalo}, {Maciel}, \&
  {Flynn}}]{rocha00}
{Rocha-Pinto}, H.~J., {Scalo}, J., {Maciel}, W.~J., \& {Flynn}, C. 2000, \aap,
  358, 869

\bibitem[{{Rosolowsky}(2007)}]{roso07}
{Rosolowsky}, E. 2007, \apj, 654, 240

\bibitem[{{Sauty} {et~al.}(1998){Sauty}, {Gerin}, \& {Casoli}}]{sauty98}
{Sauty}, S., {Gerin}, M., \& {Casoli}, F. 1998, \aap, 339, 19

\bibitem[{{Saviane} {et~al.}(2008){Saviane}, {Momany}, {da Costa}, {Rich}, \&
  {Hibbard}}]{sav08}
{Saviane}, I., {Momany}, Y., {da Costa}, G.~S., {Rich}, R.~M., \& {Hibbard},
  J.~E. 2008, \apj, 678, 179

\bibitem[{{Schechter}(1976)}]{sch76}
{Schechter}, P. 1976, \apj, 203, 297

\bibitem[{{Scheepmaker} {et~al.}(2007){Scheepmaker}, {Haas}, {Gieles},
  {Bastian}, {Larsen}, \& {Lamers}}]{scheep07}
{Scheepmaker}, R.~A., {Haas}, M.~R., {Gieles}, M., {et~al.} 2007, \aap, 469,
  925

\bibitem[{{Schweizer} {et~al.}(2008){Schweizer}, {Burns}, {Madore}, {Mager},
  {Phillips}, {Freedman}, {Boldt}, {Contreras}, {Folatelli}, {Gonz{\'a}lez},
  {Hamuy}, {Krzeminski}, {Morrell}, {Persson}, {Roth}, \&
  {Stritzinger}}]{schw08}
{Schweizer}, F., {Burns}, C.~R., {Madore}, B.~F., {et~al.} 2008, \aj, 136, 1482

\bibitem[{{Searle} \& {Zinn}(1978)}]{sz78}
{Searle}, L. \& {Zinn}, R. 1978, \apj, 225, 357

\bibitem[{{Smith} {et~al.}(2007){Smith}, {Bastian}, {Konstantopoulos},
  {Gallagher}, {Gieles}, {de Grijs}, {Larsen}, {O'Connell}, \&
  {Westmoquette}}]{smith07}
{Smith}, L.~J., {Bastian}, N., {Konstantopoulos}, I.~S., {et~al.} 2007, \apjl,
  667, L145

\bibitem[{{Smith} \& {Gallagher}(2001)}]{sg01}
{Smith}, L.~J. \& {Gallagher}, J.~S. 2001, \mnras, 326, 1027

\bibitem[{{Spitzer}(1987)}]{spit87}
{Spitzer}, L. 1987, {Dynamical evolution of globular clusters} (Princeton, NJ,
  Princeton University Press, 1987)

\bibitem[{{Spitzer}(1958)}]{spit58}
{Spitzer}, L.~J. 1958, \apj, 127, 17

\bibitem[{{Stanford} {et~al.}(1990){Stanford}, {Sargent}, {Sanders}, \&
  {Scoville}}]{stan90}
{Stanford}, S.~A., {Sargent}, A.~I., {Sanders}, D.~B., \& {Scoville}, N.~Z.
  1990, \apj, 349, 492

\bibitem[{{Terlevich}(1987)}]{terle87}
{Terlevich}, E. 1987, \mnras, 224, 193

\bibitem[{{Toomre}(1964)}]{toomre64}
{Toomre}, A. 1964, \apj, 139, 1217

\bibitem[{{van den Bergh} \& {Lafontaine}(1984)}]{vl84}
{van den Bergh}, S. \& {Lafontaine}, A. 1984, \aj, 89, 1822

\bibitem[{{Vesperini} \& {Heggie}(1997)}]{vh97}
{Vesperini}, E. \& {Heggie}, D.~C. 1997, \mnras, 289, 898

\bibitem[{{Vesperini} \& {Zepf}(2003)}]{vz03}
{Vesperini}, E. \& {Zepf}, S.~E. 2003, \apjl, 587, L97

\bibitem[{{Vogel} {et~al.}(1988){Vogel}, {Kulkarni}, \& {Scoville}}]{vogel88}
{Vogel}, S.~N., {Kulkarni}, S.~R., \& {Scoville}, N.~Z. 1988, \nat, 334, 402

\bibitem[{{Walter} {et~al.}(2002){Walter}, {Weiss}, \& {Scoville}}]{walter02}
{Walter}, F., {Weiss}, A., \& {Scoville}, N. 2002, \apjl, 580, L21

\bibitem[{{Weidner} {et~al.}(2004){Weidner}, {Kroupa}, \& {Larsen}}]{weidner04}
{Weidner}, C., {Kroupa}, P., \& {Larsen}, S.~S. 2004, \mnras, 350, 1503

\bibitem[{{Whitmore}(2003)}]{whit03}
{Whitmore}, B.~C. 2003, in A Decade of Hubble Space Telescope Science, ed.
  M.~{Livio}, K.~{Noll}, \& M.~{Stiavelli}, 153--178

\bibitem[{{Whitmore} {et~al.}(2007){Whitmore}, {Chandar}, \& {Fall}}]{whit07}
{Whitmore}, B.~C., {Chandar}, R., \& {Fall}, S.~M. 2007, \aj, 133, 1067

\bibitem[{{Wielen}(1971)}]{wielen71}
{Wielen}, R. 1971, \aap, 13, 309

\bibitem[{{Williams} \& {McKee}(1997)}]{wmk97}
{Williams}, J.~P. \& {McKee}, C.~F. 1997, \apj, 476, 166

\bibitem[{{Wilson} {et~al.}(2006){Wilson}, {Harris}, {Longden}, \&
  {Scoville}}]{wilson06}
{Wilson}, C.~D., {Harris}, W.~E., {Longden}, R., \& {Scoville}, N.~Z. 2006,
  \apj, 641, 763

\bibitem[{{Wilson} {et~al.}(2003){Wilson}, {Scoville}, {Madden}, \&
  {Charmandaris}}]{wilson03}
{Wilson}, C.~D., {Scoville}, N., {Madden}, S.~C., \& {Charmandaris}, V. 2003,
  \apj, 599, 1049

\bibitem[{{Zhang} \& {Fall}(1999)}]{zf99}
{Zhang}, Q. \& {Fall}, S.~M. 1999, \apjl, 527, L81

\bibitem[{{Zhang} {et~al.}(2001){Zhang}, {Fall}, \& {Whitmore}}]{zfw01}
{Zhang}, Q., {Fall}, S.~M., \& {Whitmore}, B.~C. 2001, \apj, 561, 727

\end{thebibliography}

\end{document}